\documentclass[12pt,preprint]{aastex}

\slugcomment{Accepted for publication in the Astronomical Journal}

\shorttitle{HST Snapshot Survey of CNOC2 Galaxy Pairs}
\shortauthors{Patton et al.}

\begin{document}

\title{A Hubble Space Telescope Snapshot Survey\altaffilmark{1} 
of Dynamically Close 
Galaxy Pairs in the CNOC2 Redshift Survey
}
\author{D. R. Patton\altaffilmark{2}, 
J. K. Grant\altaffilmark{2}, 
L. Simard\altaffilmark{3},
C. J. Pritchet\altaffilmark{4},
R. G. Carlberg\altaffilmark{5},
\& K. D. Borne\altaffilmark{6}
}

\altaffiltext{1}{
Based on observations made with the NASA/ESA Hubble Space 
Telescope, obtained at the Space Telescope Science Institute, which is 
operated by the 
Association of Universities for Research in Astronomy, Incorporated, under NASA 
contract NAS 5-26555. 
}
\altaffiltext{2}{Department of Physics, Trent University, 
1600 West Bank Drive, Peterborough, ON, K9J 7B8, Canada; dpatton@trentu.ca}
\altaffiltext{3}{Herzberg Institute of Astrophysics, National Research 
Council of Canada, 5071 West Saanich Road, Victoria, BC, V9E 2E7, Canada}
\altaffiltext{4}{Department of Physics and Astronomy, University of Victoria,
PO Box 3055 STN CSC, Victoria, BC, V8W 3P6, Canada}
\altaffiltext{5}{Department of Astronomy \& Astrophysics, 
University of Toronto, 60 St. George Street, Toronto, ON, M5S 3H8, Canada}
\altaffiltext{6}{George Mason University, School of Computational Sciences, 
MS 5c3, 4400 University Dr., Fairfax, VA 22030, USA}

\begin{abstract}
We compare the structural properties of two classes of galaxies 
at intermediate redshift: those in dynamically close galaxy pairs, and those 
which are isolated.  Both samples are selected from the 
CNOC2 Redshift Survey, and have redshifts in the range 
$0.1 < z<0.6$.  Hubble Space Telescope WFPC2 images were acquired 
as part of a snapshot survey, and were used to measure bulge fraction 
and asymmetry for these galaxies.
We find that paired and isolated galaxies have identical 
distributions of bulge fractions.  Conversely, we find that 
paired galaxies are much more likely to be asymmetric ($R_T+R_A \ge 0.13$) 
than isolated galaxies.  Assuming that half of these pairs are 
unlikely to be close enough to merge, we estimate that 
40\% $\pm$ 11\% of merging galaxies are asymmetric, compared with 
9\% $\pm$ 3\% of isolated galaxies.  The difference is even more 
striking for strongly asymmetric ($R_T+R_A \ge 0.16$) 
galaxies: 25\% $\pm$ 8\% for merging
galaxies versus 1\% $\pm$ 1\% for isolated galaxies. 
We find that strongly asymmetric paired galaxies are 
very blue, with rest-frame $B-R$ colors close to 0.80, compared
with a mean $(B-R)_0$ of 1.24 for all paired galaxies.
In addition, asymmetric galaxies in pairs have 
strong [O{\small\rm II}]3727\AA~emission lines.  We conclude that 
close to half of the galaxy pairs in our sample 
are in the process of merging, and that most of these mergers 
are accompanied by triggered star formation.

\end{abstract}

\keywords{galaxies: evolution, galaxies: interactions, galaxies: structure,
surveys}

\newpage
\section{Introduction}

Observational studies of galaxy interactions and mergers have a long
history, beginning with the identification of a sample of close galaxy 
pairs by \citet{hol37}.  Detailed multi-wavelength observations
of nearby pairs such as the Antennae (NGC 4038/4039) yield strong evidence
for ongoing mergers, including tidal tails, 
induced star formation, and multiple nuclei.  Hierarchical models of
galaxy formation indicate that mergers and accretion events 
play a key role in building up the galaxy populations we see 
today \nocite{coles05} (Coles 2005).

There is growing interest in quantifying the 
contribution that mergers have played in forming the galaxies 
we see in the nearby universe.  
Recent efforts have focussed on two central questions:
(1) how often do mergers take place? and 
(2) what effects do mergers have on the constituent galaxies?  
The answers to both questions are certain to vary with
cosmic time, and therefore require the observation of merging systems at a 
wide range of redshifts.  
In this paper, we analyze Hubble Space Telescope
(hereafter HST) images of a sample of galaxy pairs at intermediate 
redshifts, with the goal of shedding new light on both issues at a lookback 
time of several gigayears.

Attempts to address the first question have centred on 
measurement of the galaxy merger rate, and its evolution with redshift.
In recent years, many of these studies have used close galaxy pairs to 
identify systems in which mergers are imminent 
\citep{ssrs2mr,cfrs,cfgrs,cnoc2mr,bundy04,deep2}.  
The number of close companions per galaxy is taken to be 
proportional to the galaxy merger rate.
Alternatively, one can identify merging systems using 
quantitative measurements of galaxy asymmetries \nocite{con03} 
(Conselice et al. 2003 and references therein).
The resulting merger fraction
is assumed to be proportional to the galaxy merger rate.
These two methods are independent and complementary.  
Close galaxy pairs probe early stage mergers, 
while strong asymmetries are more likely to identify later stage mergers, 
including merger remnants.  There should be considerable 
overlap between these samples if the galaxy pairs are close enough
to include strongly interacting systems.

Until now, however, there has been very little evidence to confirm this 
assumption.  Most high redshift pair samples have been identified using
ground-based telescopes, and therefore yield minimal morphological 
information (e.g., Patton et al. 1997\nocite{cnoc1mr}).  
\citet{cfrs} did have HST imaging of
285 galaxies at high redshift; however, they had redshifts for only 
one member of each pair.
At low redshift, \citet{ssrs2mr} estimated
that $\sim$ 30\% of their close pairs exhibit convincing evidence of 
interactions on Digitized Sky Survey images.  
However, their classifications were
made by eye, and therefore difficult to compare with 
quantitative measurements of asymmetry.

In this study, we aim to rectify these shortcomings by analyzing 
the structural properties of dynamically close galaxy pairs identified
by Patton et al. (2002; hereafter P2002)\nocite{cnoc2mr}, 
using high resolution imaging from HST.  
The following section describes the 
observations and data reduction, and the identification of 
paired and isolated galaxies is outlined 
in \S~\ref{sample}.  A statistical comparison of the structural
parameters of
these two samples is provided in \S~\ref{comparison}.
We estimate the asymmetry fraction of paired and
isolated galaxies in \S~\ref{asymfrac}, 
and compare their star forming properties in \S~\ref{sf}.
The results are discussed in \S~\ref{discussion}, and followed
by our conclusions in \S~\ref{conclusions}.
For consistency with the sample selection of P2002,
we adopt a cosmology of
$H_0$=100 km s$^{-1}$Mpc$^{-1}$, $\Omega_M=0.2$, 
and $\Omega_{\Lambda}$=0.

\section{Observations and Data Reduction}\label{obs}

The primary goal of this study is to examine the structural properties
of galaxies in merging pairs, and to compare the results with those of
isolated galaxies.
This requires high resolution imaging of a sample of confirmed close pairs, 
along with a control sample of field galaxies.
In this section, we describe the identification of a sample of 
dynamically close galaxy pairs from the CNOC2 survey, 
and the acquisition of HST imaging for a subset of these pairs.

\subsection{Dynamical Pairs from the CNOC2 Survey}\label{cnoc2pairs}

The CNOC2 Field Galaxy Redshift Survey \citep{cnoc2} contains redshifts
for approximately 6000 galaxies at redshifts of $0.1 < z < 0.6$.  Images 
and spectra were acquired using the multi-object spectrograph (MOS)
at the Canada-France-Hawaii Telescope. 
This survey covers four well-separated patches on the sky, 
covering $1.5$ deg$^2$ to a depth of $R_C$ = 21.5, with an overall 
spectroscopic completeness of approximately 50\%.

P2002 identified a sample of 88 galaxies in dynamically close pairs
from the CNOC2 survey, and used these pairs to estimate the galaxy merger and 
accretion rates at $z=0.3$.  Galaxy pairs were identified using 
separation criteria (see \S~\ref{snapshot}), 
and without regard to morphology.  In fact, 
very little morphological detail is seen in the short 
($\sim$ 6 minute) MOS exposures, in which atmospheric seeing blurs out
most of the detail in these galaxies.  

P2002 argue that roughly half of these pairs are likely
to be undergoing mergers.  Part of the evidence comes from studying 
identically selected galaxy pairs at low redshift, using the SSRS2
Redshift Survey \citep{ssrs2mr}.  Clear morphological
signs of ongoing interactions are seen in Digitized Sky Survey
images of many of these pairs.  However, it is unclear if the same systems
are likely to be identified in deeper samples such as CNOC2, 
in part due to increased contamination by unrelated 
foreground and background galaxies.
Moreover, while the qualitative analysis of SSRS2 pairs was intriguing, 
it would be preferable to quantify morphological signs of interactions 
in an objective and reproducible manner.

\subsection{HST Snapshot Survey}\label{snapshot}

In order to address these issues, we have carried out an HST snapshot
survey of dynamically close galaxy pairs from CNOC2. 
Our full snapshot sample consisted of all CNOC2 pairs satisfying 
the following two sets of criteria.  First, all member galaxies were 
required to have
apparent magnitude $R_C \leq 21.5$, a secure spectroscopic redshift, 
and absolute magnitude $M_R^{k,e} \leq -17$, where $M_R^{k,e}$ denotes
the $k$-corrected and evolution-corrected ($E(z)$=$-z$) absolute magnitude
in the Cousins $R_C$ filter.
In addition, each pair was required to have a
projected physical separation less than $20~h^{-1}$ kpc, 
and a rest-frame relative velocity less than 500 km/s.
Approximately half of the pairs in this sample are likely to have
true physical separations of less than $20~h^{-1}$ kpc, and are
therefore expected to merge within 0.5 Gyr \citep{cnoc1mr,ssrs2mr}.

Using the preliminary CNOC2 catalogs available in advance of the HST 
observations (February 1999), this yielded a sample of 72 candidate 
galaxy pairs.  CNOC2 astrometry was used to determine the position of
the center of each galaxy pair, and the HST imaging was required to
have the pair centred on the WF3 CCD, which is relatively free of
cosmetic defects.  As each targeted pair fits easily within the 
WF3 chip, no restriction was imposed on the orientation of
WFPC2 (this maximizes the number of targets that can be observed in 
snapshot mode).
In \S~\ref{pairsample}, we discuss the final sample 
of dynamically confirmed galaxy pairs used in this study.

WFPC2 snapshot observations for this 
program (proposal ID 8230; PI Patton) were carried out
between July 1999 and June 2001.  Of the 72 galaxy pairs in 
our target list, 44 were observed.  
For each field, two 500-second F814W images were acquired, in order to
facilitate the removal of cosmic rays.  The F814W filter corresponds to
rest-frame $R$ at the mean redshift of the sample.
These images were calibrated ``on-the-fly'' as they were retrieved
from the HST archive.  The final images used in this analysis were
downloaded from the HST archive in September 2003, to ensure
consistent and updated calibrations.

\subsection{HST Image Analysis}

Each pair of images was co-added with cosmic ray rejection.  Object 
finding was then performed using SExtractor \citep{sextractor}, 
using the same source detection parameters as \citet{gim2db}.
Particular care was taken to ensure accurate de-blending of 
galaxies in close proximity to one another, while avoiding 
the detection of substructure within individual galaxies.
This was achieved by setting the SExtractor DEBLEND\_MINCONT parameter 
to 0.009.  All of the object finding was inspected by eye, and 
confirmed to be successful for all of the bright galaxies on these
images, including every galaxy used in the analysis that follows.

Additional preprocessing and image analysis was carried out using the 
Galaxy Image 2D (GIM2D) software package \citep{gim2da,gim2db}.
This IRAF/SPP package performs detailed bulge+disk decompositions of low 
signal-to-noise images of distant galaxies in a fully automated way, 
and has a proven track record with HST images \citep[e.g.,][]{tran01,im02}. 
GIM2D was used to measure a variety of structural parameters 
for each galaxy detected in our snapshot survey.  

We focus on two specific structural parameters that are best suited 
to revealing the structures of interest in this study.
The first is the bulge fraction ($B/T$), which describes
the overall structure of each galaxy.
This parameter is a measure of the fraction of the galaxy light
attributed to the bulge component.  This quantity is determined using 
bulge+disk decomposition.  A pure disk system has $B/T$=0, and a 
pure bulge system has $B/T$=1.

The second structural parameter we employ is asymmetry, 
which is used to detect morphological signs of interactions 
that may be present.
While GIM2D makes several measurements of asymmetry, we choose to use 
$R_T+R_A$, which is a measure of the overall smoothness of the galaxy image 
with respect to the bulge+disk model. 
$R_T$ and $R_A$ are defined in equation 11 of \citet{gim2db}, and 
are the GIM2D versions of the parameters originally 
introduced by \citet{schade95}.  
$R_T$ and $R_A$ are approximately equal to the 
total and asymmetric residual flux respectively, 
and are expressed as a fraction of the total galaxy flux.  
The discriminating power of $R_T+R_A$, as computed by 
GIM2D, has been demonstrated by \citet{im02} and \citet{mcintosh04}.
We compute $R_T+R_A$ within two half light radii.  This region is 
large enough to sample the vast majority of the asymmetric light 
in these systems.  Using a larger aperture could, in principle, 
increase contamination due to light from neighbouring galaxies.
However, we note that the main conclusions of this study are unchanged 
by measuring asymmetry within three, four, or five half light radii.
We conclude that this asymmetry parameter should be sensitive to 
the most significant morphological signs of interactions that are present 
in this sample.

\section{Extraction of Paired and Isolated Galaxies}\label{sample}

\subsection{Paired Galaxy Sample}\label{pairsample}

In this section, we describe the identification of a secure sample
of paired galaxies found on the HST images.  
All 44 images were inspected visually to confirm the presence of
a galaxy pair at the expected location (the centre of the WF3 CCD).
In one instance, a single galaxy was found (Target 64).  
In this case,
it appears that the ground-based CNOC2 survey mistakenly identified
a bright region within a galaxy as a separate galaxy.  In addition, 
revised redshift measurements in the final CNOC2 catalog led to the 
removal of two pairs (Targets 03 and 71) from the sample, as their 
member galaxies are no longer found to be at similar redshifts.
Three pairs (Targets 22, 23, and 43) were known in advance to have only a 
single redshift, and hence were discarded 
from this pairs sample.  
Finally, given that pairs with widely differing masses are less likely 
to exhibit interaction-induced structural changes, a restriction of 
2.5 magnitudes difference in absolute magnitude was imposed.
This removed three additional pairs (Targets 02, 62, and 63) 
from the sample.  

We are left with a sample of 35 dynamically close galaxies pairs
in these HST images, containing 70 paired galaxies.  
We note that our definition of a close companion allows for triples 
and higher order groupings; however, the original snapshot sample included
only one triple, and this system was not observed with HST.
The basic properties of the pairs are listed in Table~\ref{tabpairprop}, 
including: HST Target number, mean redshift, 
angular separation ($\theta$), projected physical separation ($r_p$), 
rest-frame relative velocity ($\Delta v$), and difference in 
absolute magnitude ($\Delta M_R$).  We note that some of these pairs have
slightly different projected separations than reported by P2002.
due to improved angular pair separation measurements from HST imaging.

\begin{figure}
\epsscale{0.9}
\plotone{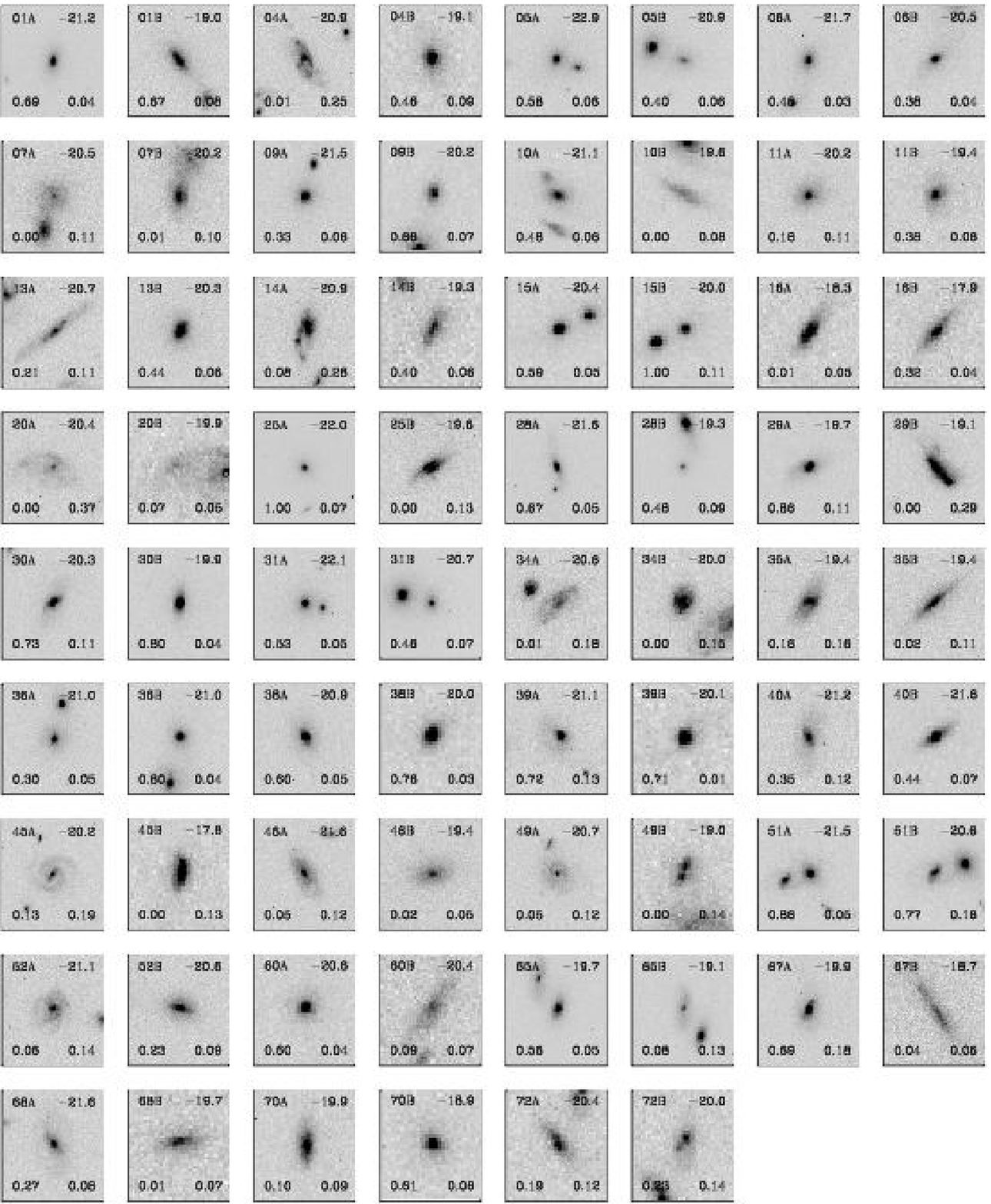}
\caption{Mosaic of the 70 paired galaxies listed in Table~\ref{tabpaired}.
Each image is centred on a paired galaxy, and is labelled with:
HST ID (upper left), absolute magnitude $M_R$ (upper right), 
bulge fraction $B/T$ (lower left), and asymmetry $R_T+R_A$ (lower right).
The extracted region for each galaxy is 20 times the area of its 
1.5-$\sigma_{\rm bkg}$ isophote.  
No orientation was specified for these snapshot observations; 
images therefore have random orientation angles.
\label{paired}}
\end{figure}

WFPC2 F814W postage stamp images of the galaxies in these pairs 
are presented in Figure~\ref{paired}.
Table~\ref{tabpaired} contain properties of these 
paired galaxies, including: HST ID (HST Target; galaxy A or B), 
CNOC2 name (patch and serial number), HST astrometry 
(J2000), spectroscopic redshift, 
apparent magnitude in Cousins $R$, and absolute magnitude in 
Cousins $R$ ($M_R$).
Additional properties given in these tables are 
described later in the paper.

\subsection{Control Sample}\label{control}

While this snapshot survey targeted galaxy pairs, 
many other CNOC2 galaxies are found on these WFPC2 images.
This allows for the identification of a control sample of 
field galaxies at similar redshifts.
This is an {\it essential} component of this analysis, as it enables
us to determine if paired galaxies differ significantly from 
isolated field galaxies.

We require the control sample to be chosen in 
precisely the same manner as paired galaxies, with the exception
of the requirement of a dynamically close neighbour.  Of greatest
importance is the availability of a spectroscopic redshift from CNOC2, 
so as to ensure a fair comparison with the pair sample, and to confirm
isolation.  We then apply identical apparent and absolute magnitude cuts as for 
the paired galaxies: namely, $R_C < 21.5$ and $M_R^{k,e} \leq -17$.
These redshift catalogs were cross-correlated with the HST images, 
in order to identify candidates for the control sample.  Any galaxies 
lying too close to the edges of the WF chips were excluded from the analysis.
In addition, a number of galaxy images were adversely affected by
bad columns on the CCDs, preventing accurate measurement of 
structural parameters; hence they too were removed from the control sample
(note that this did not happen with any of the paired galaxies, 
since all pairs were centred on a clean region in the center of the WF3 chip).
This yielded a control sample of 157 field galaxies.

\subsection{Isolated Galaxy Sample}

As stated earlier, the primary goal of this study is to determine
what effect (if any) the presence of a close companion has on 
a paired galaxy.  To this end, we would ideally like to compare 
the paired galaxy sample with field galaxies which are certain 
to have no close companions.  A cursory inspection of the control
sample reveals that in fact many of these galaxies do appear to 
have relatively nearby companions.  For approximately half of 
these companions, we can use their redshifts to determine if 
they are at a similar redshift or not.  For the remainder, 
the lack of redshift information means that we cannot rule them out
as possible physical companions.

\begin{figure}
\epsscale{0.9}
\plotone{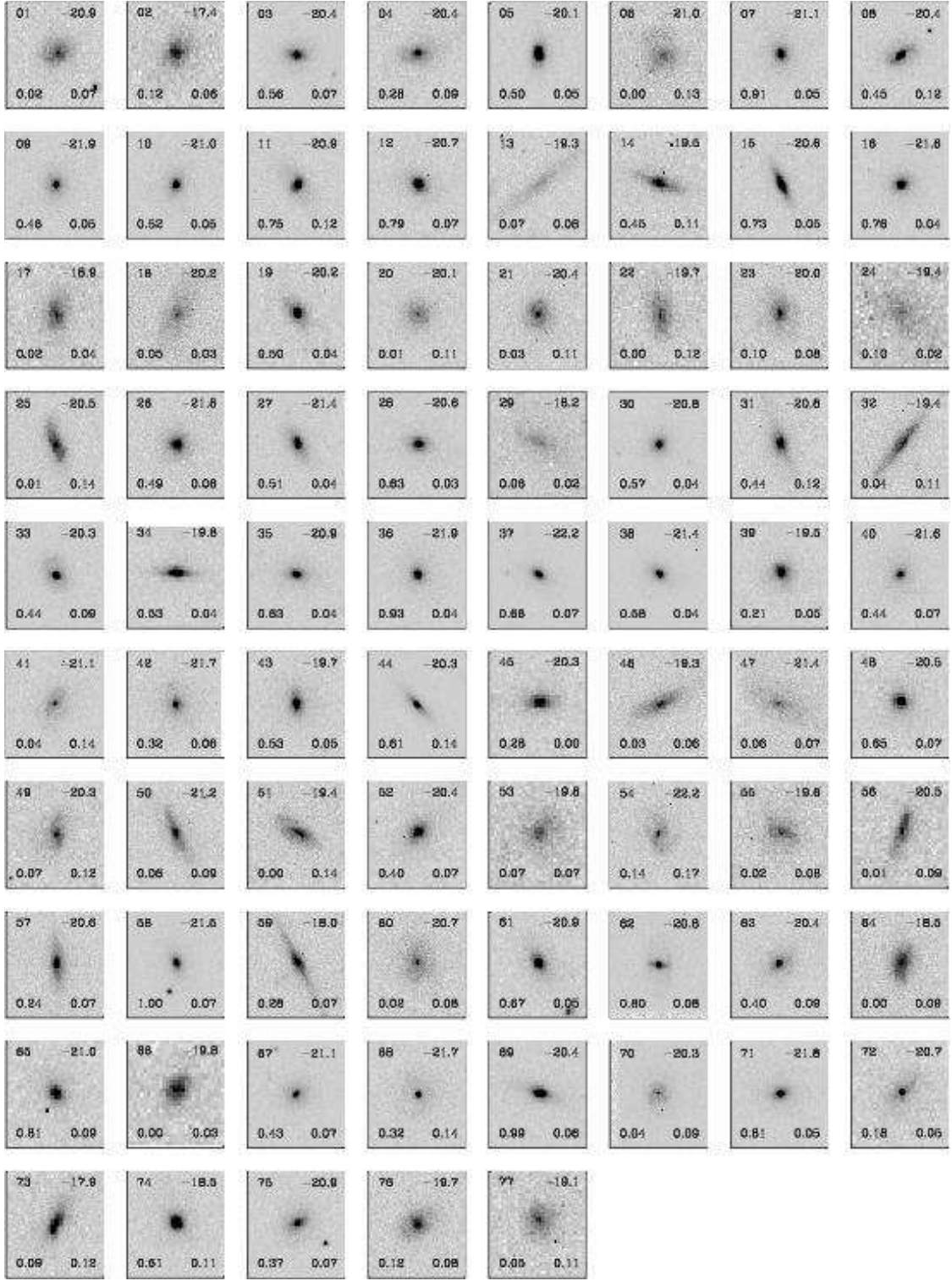}
\caption{Mosaic of the 77 isolated galaxies listed in Table~\ref{tabiso}.
Each image is centred on an isolated galaxy, and has the same labelling and
display format as in the preceding figure.
\label{iso}}
\end{figure}

With this in mind, we apply several additional criteria to the control 
sample to ensure that all remaining galaxies are truly isolated.
A neighbouring galaxy is initially considered to be a companion if it 
lies within a projected separation of 40 $h^{-1}$ kpc and has an
observed flux larger than 20\% (ie., within 1.75 magnitudes) 
of the host galaxy.  Any companion with a rest-frame relative velocity 
larger than 1000 km/s is discarded.
Every galaxy in the control sample which has no companions is 
then classified as isolated.  
These criteria yield a sample of 77 isolated galaxies.
Postage stamp images of these isolated galaxies are shown in 
Figure~\ref{iso}.  Table~\ref{tabiso} contains a list of properties 
of these galaxies, in the same format as Table~\ref{tabpaired}.

\subsection{Redshift Distributions}

\begin{figure}
\plotone{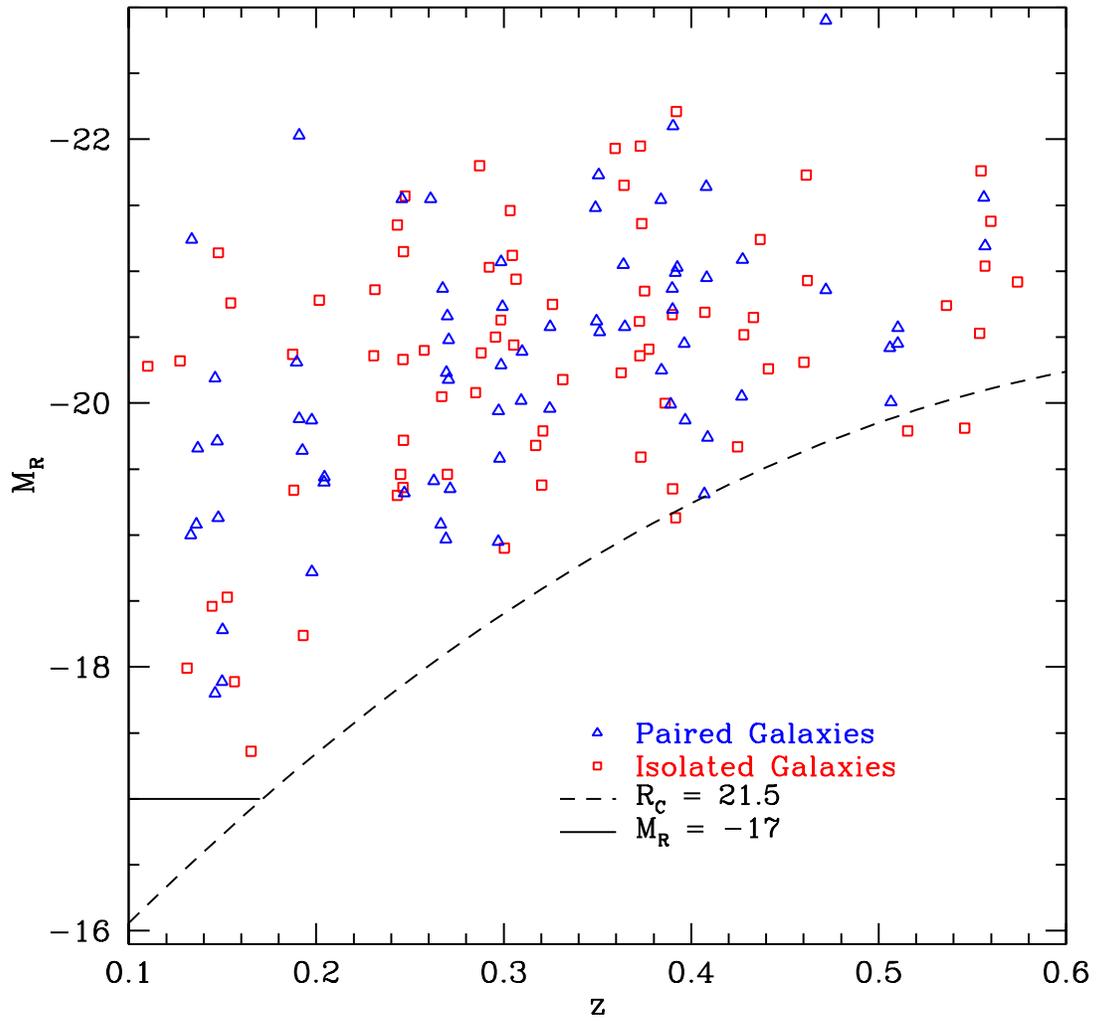}
\caption{Absolute magnitude is plotted versus redshift for 
paired galaxies (blue triangles) and isolated galaxies (red squares).
While the overall distributions of the two samples are similar, 
isolated galaxies are found at slightly higher redshifts, as expected.
\label{absmag}}
\end{figure}

The redshift-absolute magnitude distributions of paired and isolated
galaxies are compared in Figure~\ref{absmag}.  
Overall, these samples are seen to have similar distributions.  
However, isolated galaxies appear to 
lie at slightly higher redshifts than paired galaxies.
This is a natural consequence of the selection criteria imposed.
Optical contamination due to higher redshift interlopers 
is expected to be higher for galaxies
at low redshift, due to the larger angular search area.  As a result,
it is more difficult to confirm isolation for galaxies at the low redshift
end of the sample.  In addition, the detection of 
galaxy pairs depends on the square of the selection function, 
meaning that pairs will be relatively rare at the high redshift
end of the sample \citep{ssrs2mr}.  

However, the difference in mean redshift between paired and isolated 
galaxies is small (0.03), and is significant only at the 1$\sigma$ level.
Consequently, differences in the redshift distributions of paired and 
isolated galaxies are unlikely to cause a statistically significant 
difference in the structural parameters of these samples.
Therefore, we ignore this effect throughout the remainder of this paper.

\section{Structural Parameters of Paired and Isolated Galaxies}\label{comparison}

The primary goal of this study is to measure the structural 
parameters of galaxies in dynamically close galaxy pairs, 
and to compare these measurements with those of a 
sample of isolated field galaxies.  
In the sections that follow, we compare the distributions of 
$B/T$ and $R_T+R_A$ of paired and isolated galaxies.
Table~\ref{tabprop} lists these and other statistical properties 
of paired and isolated galaxies, including the results of
KS tests comparing these two samples.  

\subsection{Bulge Fraction}\label{bulgefraction}

\begin{figure}
\epsscale{1.0}
\plotone{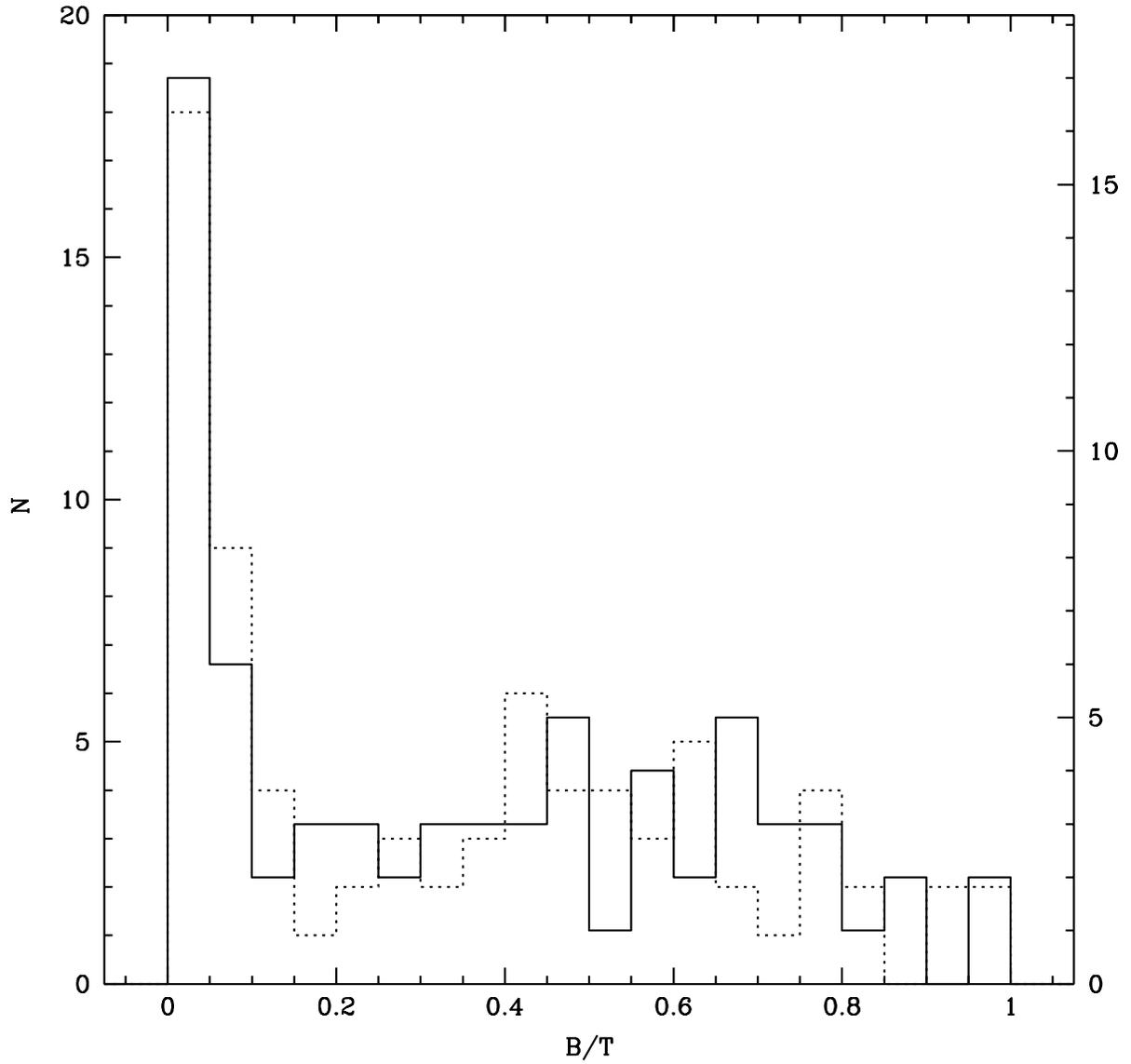}
\caption{Histograms of $B/T$ are given for 
paired galaxies (solid line) 
and isolated galaxies (dotted line).
The vertical axis on the left gives the number of isolated galaxies, 
and the vertical axis on the right gives the number of paired galaxies.
The histograms have been normalized such that the area under each 
histogram is the same.  No obvious differences are seen in the 
two distributions.
\label{fighistbf}}
\end{figure}

A histogram of $B/T$ for paired and isolated galaxies is 
presented in Figure~\ref{fighistbf}.  It is immediately 
apparent that these samples have very similar distributions.  
We carry out two tests to verify this statistically.
First, we compute the mean of each distribution, 
along with the error in the mean (see Table~\ref{tabprop}).
The means are found to be statistically identical.  
Next, we perform a Kolmogorov-Smirnov (K-S) test on the 
two distributions, in which the K-S probability indicates 
the significance level for the hypothesis
that two samples are drawn from the same distribution.
A small significance level indicates that two distributions
have significantly different cumulative distribution functions.
For the $B/T$ distributions of paired and isolated galaxies, 
we find a KS probability of 95\%, indicating that we do not 
detect any statistical difference in the samples.  
Therefore, we conclude that paired and isolated galaxies have statistically 
equivalent distributions of $B/T$. 

\subsection{Asymmetry}\label{asym}

\begin{figure}
\epsscale{1.0}
\plotone{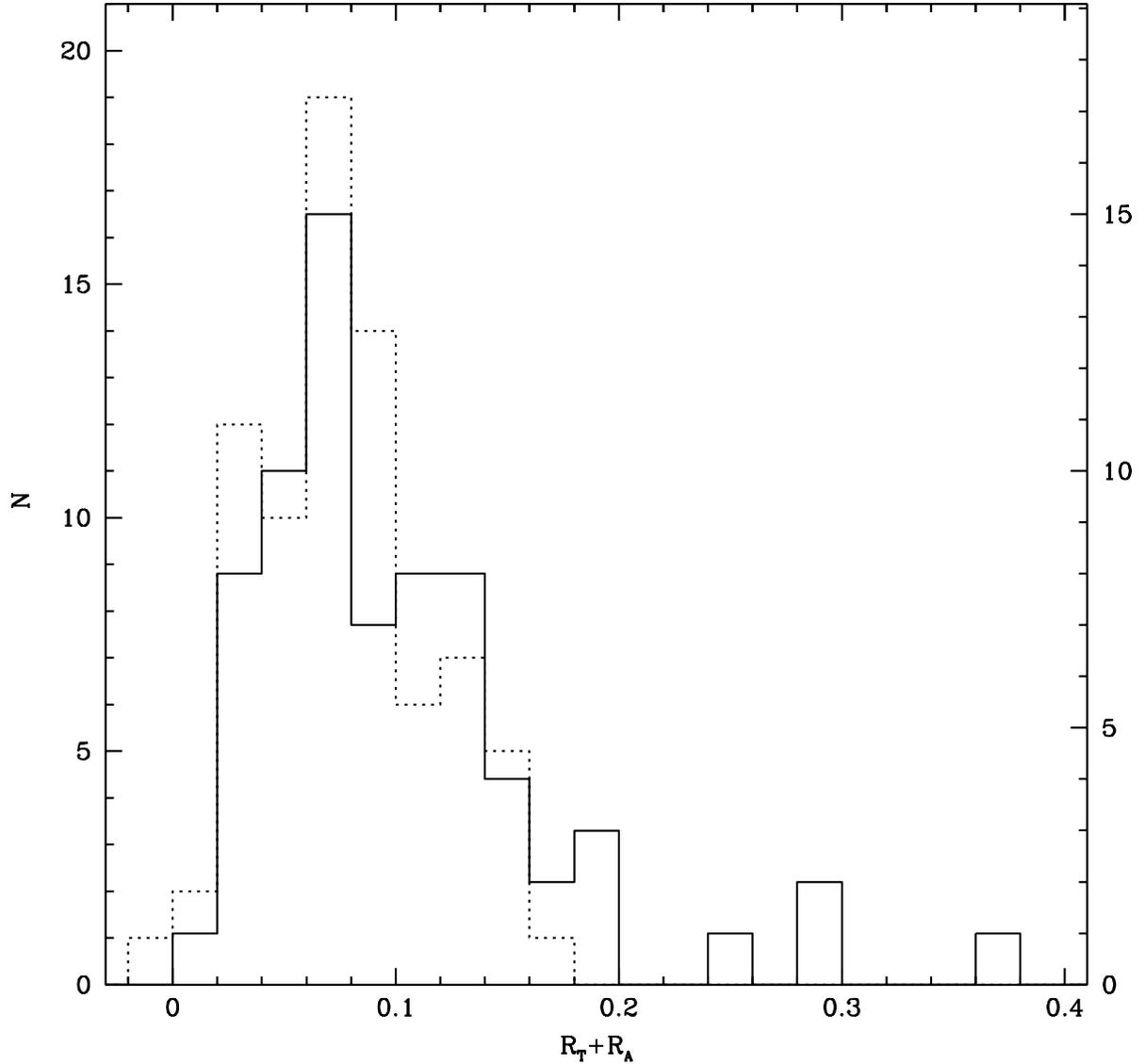}
\caption{Histograms of $R_T+R_A$ are given for 
paired galaxies (solid line) and isolated galaxies (dotted line).
The vertical axis on the left gives the number of isolated galaxies, 
and the vertical axis on the right gives the number of paired galaxies.
The histograms have been normalized such that the area under each 
histogram is the same.  The high asymmetry regime is dominated by
paired galaxies.
\label{fighista}}
\end{figure}

A histogram of $R_T+R_A$ for paired and isolated galaxies is 
presented in Figure~\ref{fighista}.  These distributions 
appear to differ substantially, particularly in the 
high asymmetry regime, where a noticeable tail is seen in the distribution
of paired galaxies.
This is precisely the regime in which one might expect interacting 
galaxies to be more prevalent.  Table~\ref{tabprop} shows 
that the mean $R_T+R_A$ of paired galaxies 
is larger than that of isolated galaxies, 
and the difference is significant at the 1.8$\sigma$ level.
The results of the K-S test confirm this, with a K-S probability of only
16\%.  We conclude that a subset of the paired galaxies are significantly 
more asymmetric than isolated galaxies.  

\subsection{Bulge Fraction versus Asymmetry}

\begin{figure}
\epsscale{1.0}
\plotone{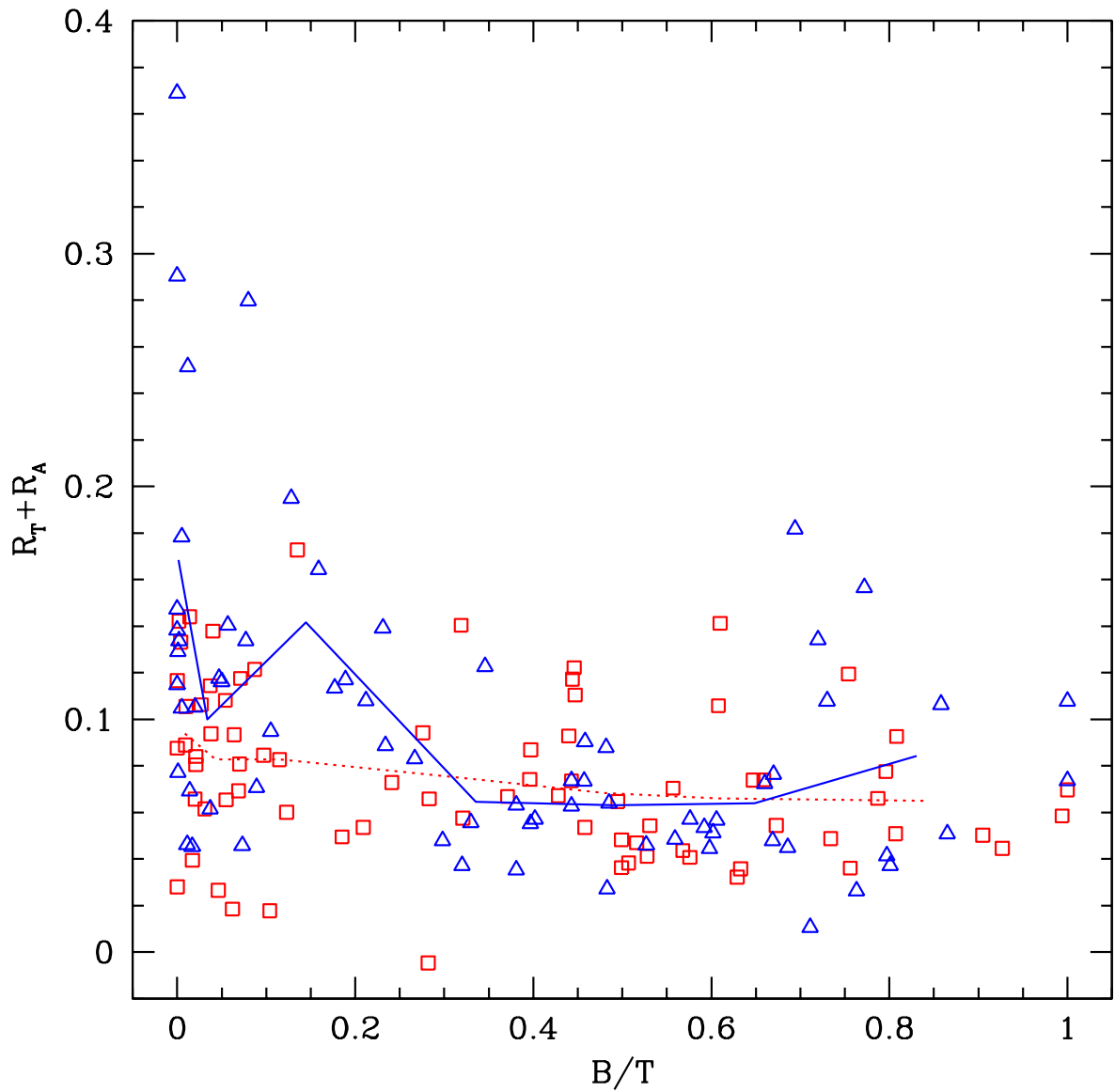}
\caption{Asymmetry is plotted versus bulge fraction for paired galaxies 
(blue triangles) and isolated galaxies (red squares).
The dotted red line gives the mean asymmetry of isolated 
galaxies in seven bins of $B/T$, and shows a small decrease
of asymmetry with increasing bulge fraction.  The solid blue
line gives the mean asymmetry of paired galaxies in seven bins of $B/T$. 
The most obvious difference
between the samples is seen at low bulge fractions, where the mean 
asymmetry of paired galaxies is noticeably higher than for isolated galaxies.
\label{figbfrtra1}}
\end{figure}

In order to better understand these correlations, 
we investigate the relationship 
between bulge fraction and asymmetry for paired and isolated 
galaxies.  Figure~\ref{figbfrtra1} exhibits several interesting 
trends.  First, for isolated galaxies, there is a small monotonic decrease 
in asymmetry as bulge fraction increases.  The locus of this distribution 
is similar to that found using galaxies from the Sloan Digital 
Sky Survey \citep{sdssdr2}, 
using identical GIM2D parameters (Simard 2005, in preparation).

This trend is also seen for paired galaxies, though there is much
more scatter in the distribution of asymmetries.
Overall, paired galaxies appear to have larger asymmetries 
than isolated galaxies, particularly for low bulge fractions.
The most asymmetric paired galaxies ($R_T+R_A > 0.2$) 
all have low bulge fractions ($B/T < 0.2$).
We will investigate the significance of these trends further
in \S~\ref{asymfrac}.

\subsection{Potential Bias Due to Crowding of Galaxies}

One possible explanation for the differences seen between paired
and isolated galaxies is a bias in the measurement of structural 
parameters for galaxies which are crowded closely together on the sky.
That is, in attempting to measure structural parameters (especially 
asymmetry) of a given galaxy, it is possible that the measurements
may be adversely affected by the presence of a nearby galaxy.
In particular, asymmetry measurements could be compromised if the light
from a neighbouring galaxy contributes unequal and significant amounts 
of light to opposite sides of the galaxy in question.

\begin{figure}
\epsscale{1.0}
\plotone{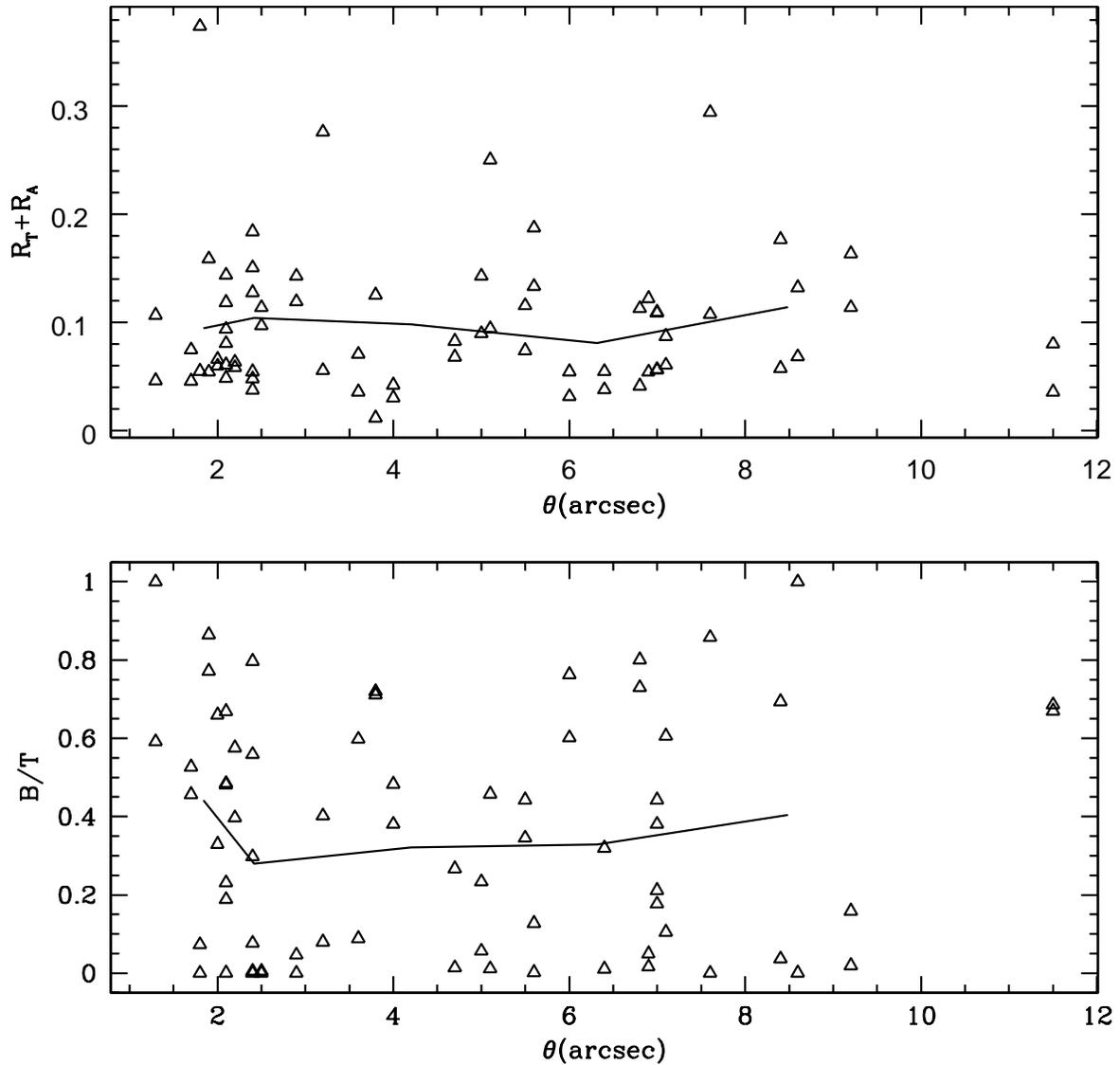}
\caption{Bulge fraction and asymmetry of paired galaxies are
plotted versus pair angular separation $\theta$.
The solid lines give the mean bulge fraction and asymmetry 
in five bins of $\theta$ (14 galaxies per bin).
No significant dependence on $\theta$ is seen.  This supports the 
assumption that nearby companions do not cause a significant bias
in the measurement of structural parameters.
\label{figtheta}}
\end{figure}

If true, we would expect to see significant trends in observed 
structural parameters with angular separation of the two galaxies.
If this bias is completely responsible for the observed differences
between paired and isolated galaxies, we would expect the anomalous 
paired galaxies to be found preferentially at small angular separations.
To investigate this possibility, we plot the structural 
parameters of paired galaxies versus the angular separation, 
in Figure~\ref{figtheta}.  
No significant dependence on $\theta$ is seen in this figure.  
In particular, galaxies with large asymmetries and large bulge fractions 
are seen at a wide range of angular separations.  This supports the 
assumption that nearby companions do not cause a significant bias
in the measurement of structural parameters.  Therefore, we ignore
this effect throughout the remainder of this paper. 

\subsection{Dependence on Pair Separation and Luminosity Ratio}

\begin{figure}
\epsscale{1.0}
\plotone{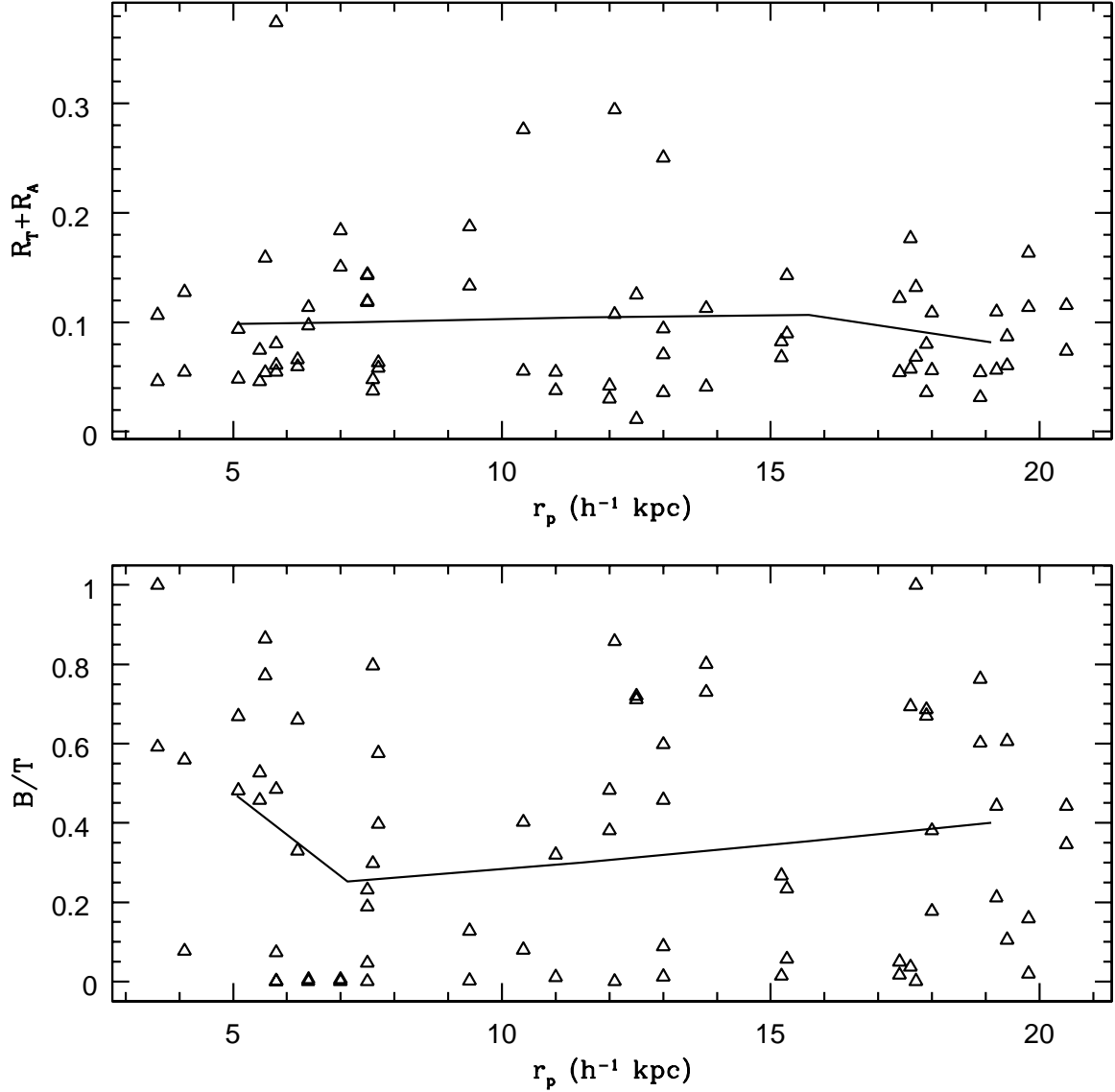}
\caption{Bulge fraction and asymmetry of paired galaxies are
plotted versus pair projected physical separation $r_p$.
The solid lines give the mean bulge fraction and asymmetry 
in five bins of $r_p$.
No significant dependence on $r_p$ is seen. 
\label{figrp}}
\end{figure}

\begin{figure}
\epsscale{1.0}
\plotone{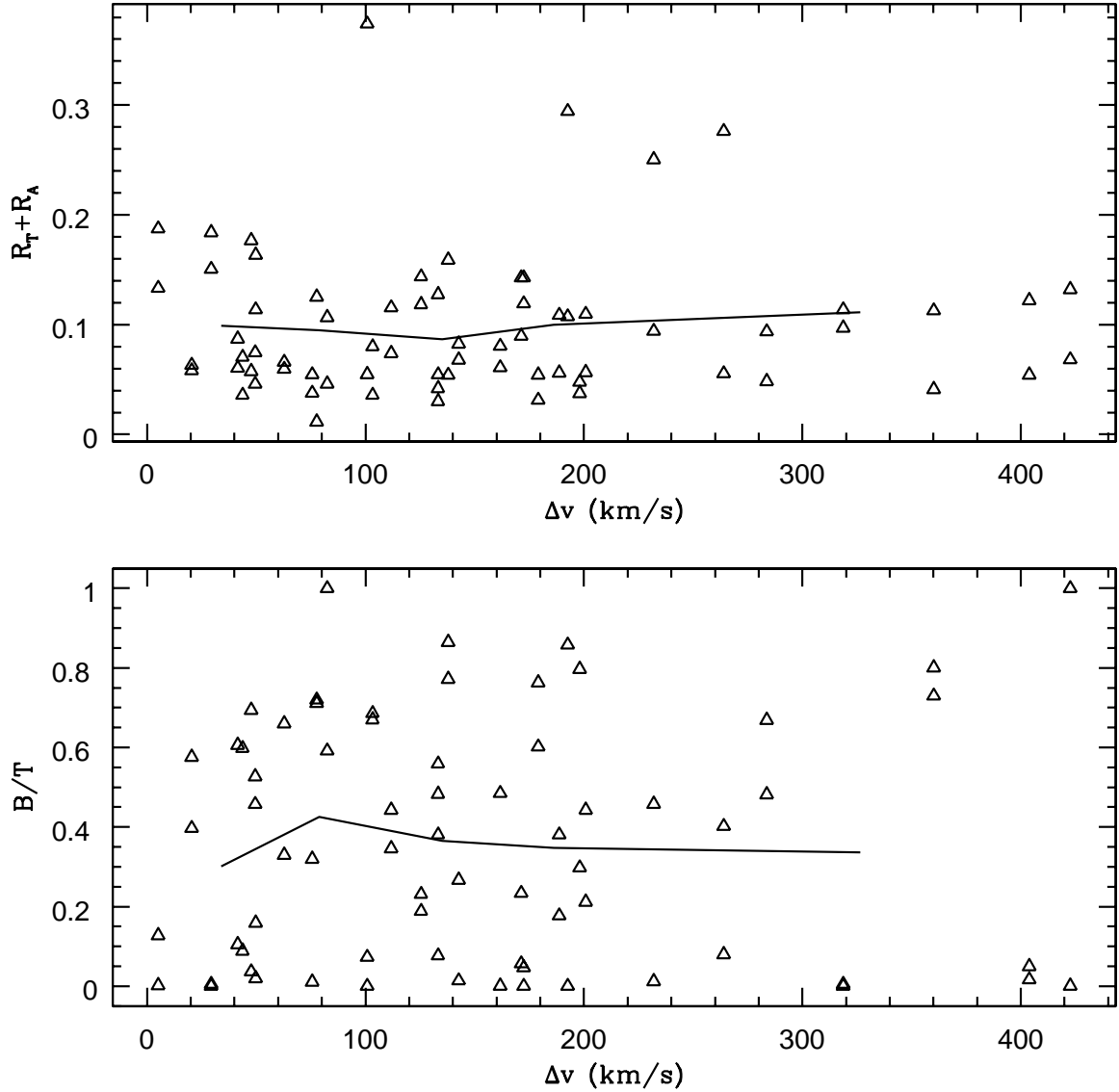}
\caption{Bulge fraction and asymmetry of paired galaxies are
plotted versus pair rest-frame relative velocity $\Delta v$. 
The solid lines give the mean bulge fraction and asymmetry 
in five bins of $\Delta v$.
No significant dependence on $\Delta v$ is seen. 
\label{figdelv}}
\end{figure}

Our sample of galaxy pairs spans a small range in both 
projected separation ($r_p$) and line of sight velocity 
difference ($\Delta v$).  Nevertheless, we now explore the possibility 
that the structural parameters depend on this separation.
Measured structural parameters are plotted versus 
$r_p$ and $\Delta v$ in Figures~\ref{figrp} and \ref{figdelv}
respectively.  No obvious trend is observed.  In particular, galaxies with 
relative large asymmetries and bulge fractions are found over the 
full range of parameter space explored: namely, 
$3~h^{-1}$ kpc $< r_p < 21~h^{-1}$ kpc and $\Delta v < 425$ km/s.
This is not particularly surprising, given the limited sample size
and the relatively small range of pair separations available.  
Moreover, the use of projected separation (rather than three-dimensional 
separation) weakens any existing correlation between pair separation and 
galaxy properties.  With a 
sample spanning greater pair separations, however, one might well expect to
see a stronger dependence of structural properties on pair separation 
(e.g., \citealt{her05}).  

\begin{figure}
\epsscale{1.0}
\plotone{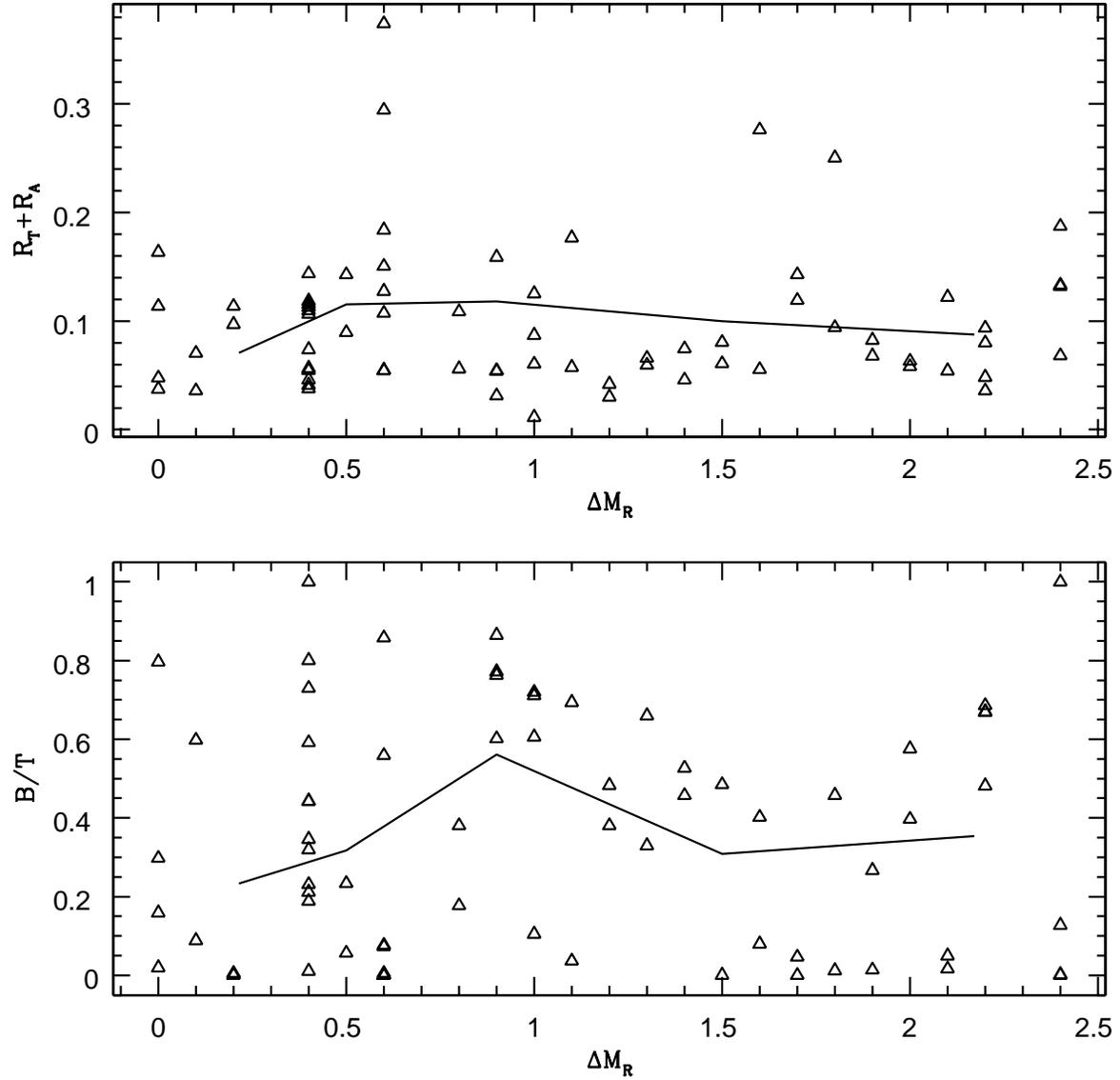}
\caption{Bulge fraction and asymmetry of paired galaxies are
plotted versus pair absolute magnitude difference $\Delta M_R$.
The solid lines give the mean bulge fraction and asymmetry 
in five bins of $\Delta M_R$.
No significant dependence on $\Delta M_R$ is seen. 
\label{figdelm}}
\end{figure}

In addition, one might expect structural properties to depend on 
the relative mass of the galaxies in merging pairs.  In particular, 
equal mass pairs (major mergers) might be expected to exhibit stronger
signs of interactions than pairs with significantly difference masses 
(minor mergers).
We use the difference in absolute magnitude $M_R^{k,e}$ (hereafter 
$\Delta M_R$) as our best indicator of this ratio.  The maximum 
$\Delta M_R$ in our sample is 2.5 (see \S~\ref{pairsample}), corresponding to 
a 1:10 ratio in luminosity.  
Structural properties are plotted versus $\Delta M_R$ 
in Figure~\ref{figdelm}.  Once again, no strong correlation is seen.
This indicates that any merger-induced changes in 
morphology are seen throughout the full range of mass ratios probed 
by our sample.

\section{The Asymmetry Fraction}\label{asymfrac}

One of the primary goals of this study is to determine whether, and 
how often, paired galaxies exhibit morphological signs of interactions.  
To this end, we investigate the fraction of galaxies with 
measured asymmetry in excess of a given threshold.  
This approach has been used by \citet{con03} to estimate the
fraction of galaxies undergoing mergers.  

We begin by selecting an asymmetry threshold that is likely to
distinguish between normal galaxies and those with morphological
signs of interactions.  Despite the objective and quantitative 
nature of these asymmetry measurements, the choice of a threshold
is somewhat subjective.  Images of all 
of the galaxies in both the paired and isolated samples were 
inspected by eye, to ascertain whether or not the asymmetries 
seen were substantial. 
A threshold of $R_T+R_A = 0.13$ was found to be the best threshold 
for separating symmetric and asymmetric galaxies.  
Hereafter, we refer to all galaxies which have $R_T+R_A \ge 0.13$ 
as being ``asymmetric''.  
This is slightly less strict than 
the threshold of $R_T+R_A > 0.14$ used by \citet{schade95}.

\begin{figure}
\epsscale{1.0}
\plotone{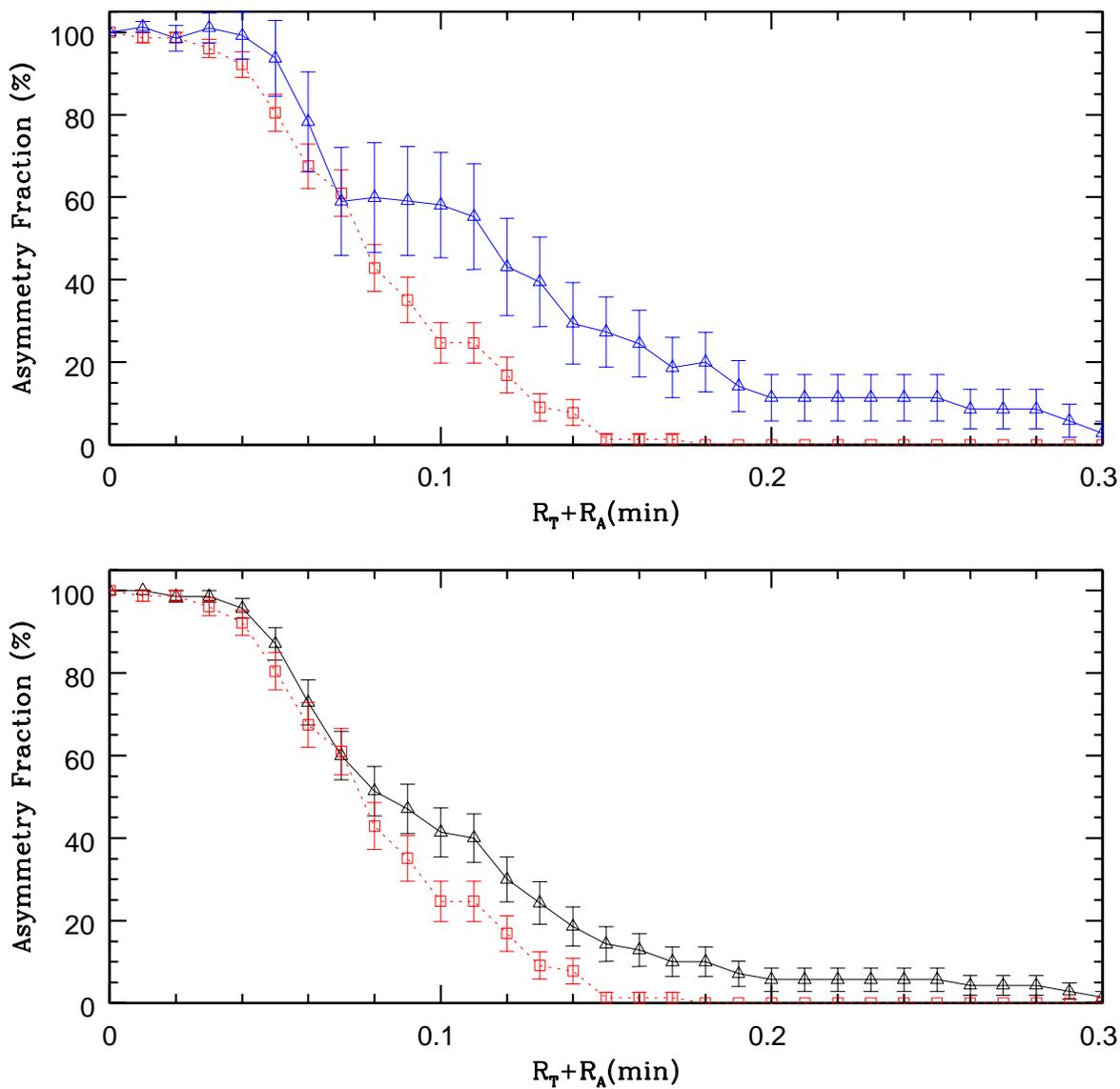}
\caption{
The fraction of galaxies more asymmetric than $R_T+R_A({\rm min})$ is 
plotted versus $R_T+R_A({\rm min})$.  
(a) The lower plot compares paired galaxies 
(black triangles connected with a solid line) with isolated galaxies 
(red squares connected with a dotted line).  (b) The upper plot 
compares ``merging'' galaxies (blue triangles connected with a solid line) 
with isolated galaxies (same symbols as lower plot).  All error bars are
one sigma.
\label{figfpfi}}
\end{figure}

Using this definition, we find that $24.3 \pm 5.2\%$ of paired galaxies
are asymmetric, compared with $9.1 \pm 3.3\%$ of isolated galaxies. 
That is, we find that paired galaxies are nearly three times more likely 
to be classified as asymmetric, and the difference between these samples
is significant at the 1.8$\sigma$ level.  A comparison of asymmetry 
fractions using other choices of thresholds can be seen in the lower 
panel of Figure~\ref{figfpfi}.  This plot reveals a statistically 
significant difference between the asymmetry fractions of 
paired and isolated galaxies for a wide range of asymmetry thresholds, 
strengthening this conclusion.

Moreover, it is important to note that not all galaxies in the paired sample
are likely to be undergoing interactions or mergers.  Instead, the 
separation criteria used to identify galaxy pairs are less than perfect at 
choosing merging pairs, due mainly to the contribution of peculiar
velocities to the line-of-sight separation of the galaxies in 
a given pair.  As discussed by
\citet{ssrs2mr}, some of the dynamical galaxy pairs 
identified with these criteria are likely to have true physical 
separations that are so large that the system will never merge.
The fraction of pairs that are likely to have true 
three-dimensional physical separations of less than 20 $h^{-1}$ kpc
has been estimated to be $f_{\rm 3D} = 0.5$ 
\citep{ssrs2mr,cnoc2mr}.   
While this estimate is approximate in nature, it is greatly 
preferable to assuming that all of the paired galaxies in 
our study are close enough to merge.

With this assumption, we infer that our sample of paired galaxies 
is made up of equal parts isolated galaxies and merging galaxies.
Using this approach, we find that $39.5 \pm 10.9\%$ of merging galaxies are
asymmetric.  This is 4.3 times larger than the asymmetry fraction of isolated 
galaxies.
In the upper panel of Figure~\ref{figfpfi}, 
the asymmetry fraction of merging and isolated galaxies is plotted
over a wide range in asymmetry thresholds.  This figure shows that
the difference between the two samples is striking, and is not 
confined to our specific choice of asymmetry threshold.  
We note also that this result has implications for the K-S test 
applied to the paired and isolated samples in \S~\ref{asym}; 
that is, the K-S probability of 16\% is almost certainly too high, 
due to the mixture of merging and isolated galaxies in the paired sample.  
We conclude that merging galaxies are significantly more asymmetric
than isolated galaxies.

Finally, we repeat this analysis using a stricter asymmetry threshold.
Visual inspection of the images indicates that galaxies 
with $R_T+R_A \ge 0.16$ may be described as ``strongly asymmetric''.
With this definition, we find that $1.3 \pm 1.3\%$ of isolated galaxies
are strongly asymmetric, $12.9 \pm  4.0\%$ of paired galaxies are 
strongly asymmetric, 
and $24.5 \pm 8.1\%$ of merging galaxies are
strongly asymmetric.  Clearly, the difference between the samples is 
even more striking when the most strongly asymmetric galaxies 
are considered.  

\section{Colors and Emission Line Strengths}\label{sf}

\subsection{Rest-Frame Colors}

Historically, there have been many attempts to detect enhanced 
star formation in merging galaxies (e.g., Larson \& Tinsley 1978)\nocite{LT}.
Optical colors of galaxies are known to be useful, though imperfect, indicators
of their underlying stellar populations and star-forming histories. 
The HST imaging used in this study are in a single passband (F814W) only; 
however, we do have integrated 
color measurements from the CNOC2 survey \citep{cnoc2}.  
Rest-frame colors were computed by fitting model spectral energy 
distributions to the CNOC2 photometry ($UBVR_CI_C$).  
We limit our analysis to rest-frame $B-R$ color, hereafter $(B-R)_0$.
The mean colors and cumulative distributions of paired and isolated galaxies
are found to be statistically equivalent (see Table~\ref{tabprop}). 

\begin{figure}
\epsscale{1.0}
\plotone{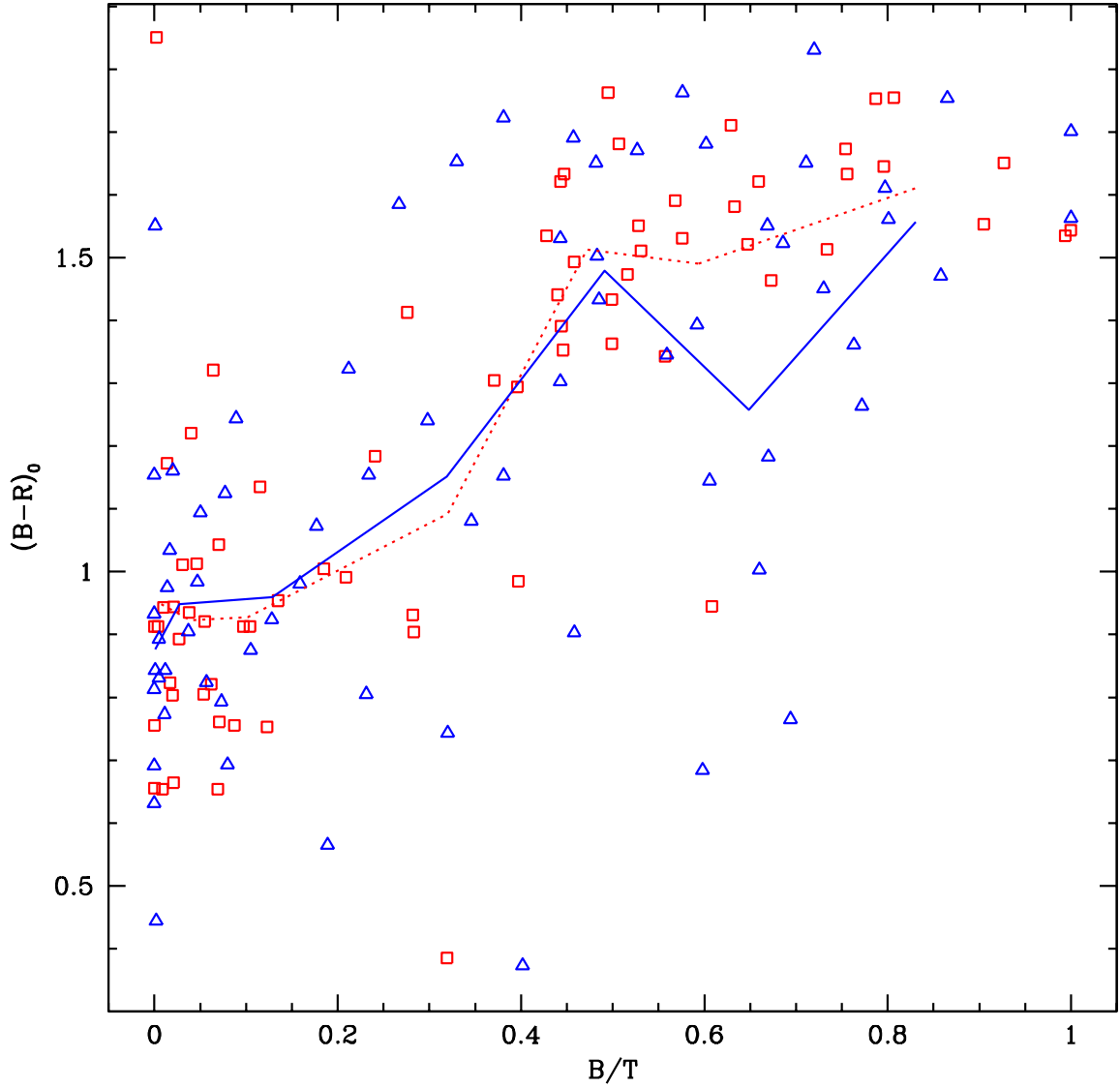}
\caption{Rest-frame $B-R$ color is plotted versus bulge fraction 
for paired (blue triangles) and 
isolated (red squares) galaxies.
The dotted red line gives the mean color of isolated 
galaxies in seven bins of $B/T$, and shows a clear correlation between
color and bulge fraction.
Similarly, the solid blue line gives the mean color of paired galaxies.
\label{figbr0bf}}
\end{figure}

We now investigate the connection between rest-frame colors 
and structural parameters.  Figure~\ref{figbr0bf} contains a plot
of rest-frame color versus bulge fraction, for paired and isolated galaxies.
A clear correlation is seen for isolated galaxies, in that galaxies with 
larger bulge fractions are redder.  This result 
is consistent with numerous studies of the properties of 
field galaxies \citep[e.g.,][]{roberts}.
A similar correlation is seen for paired galaxies, 
though there is significantly more scatter in color.
Disk-dominated ($B/T < 0.5$) paired galaxies have the same 
mean color as their isolated counterparts, whereas bulge-dominated
paired galaxies are bluer (at the 1.3$\sigma$ level).
If we instead treat bulge fraction as the dependent variable, 
we find that the bluest paired galaxies ($(B-R)_0 < 1$) 
have higher bulge fractions than the bluest isolated galaxies 
(significant at 1.5$\sigma$).  These results are consistent with a 
scenario in which mergers have induced central starbursts in some paired
galaxies, making them bluer and/or boosting their measured bulge fractions. 

\begin{figure}
\epsscale{1.0}
\plotone{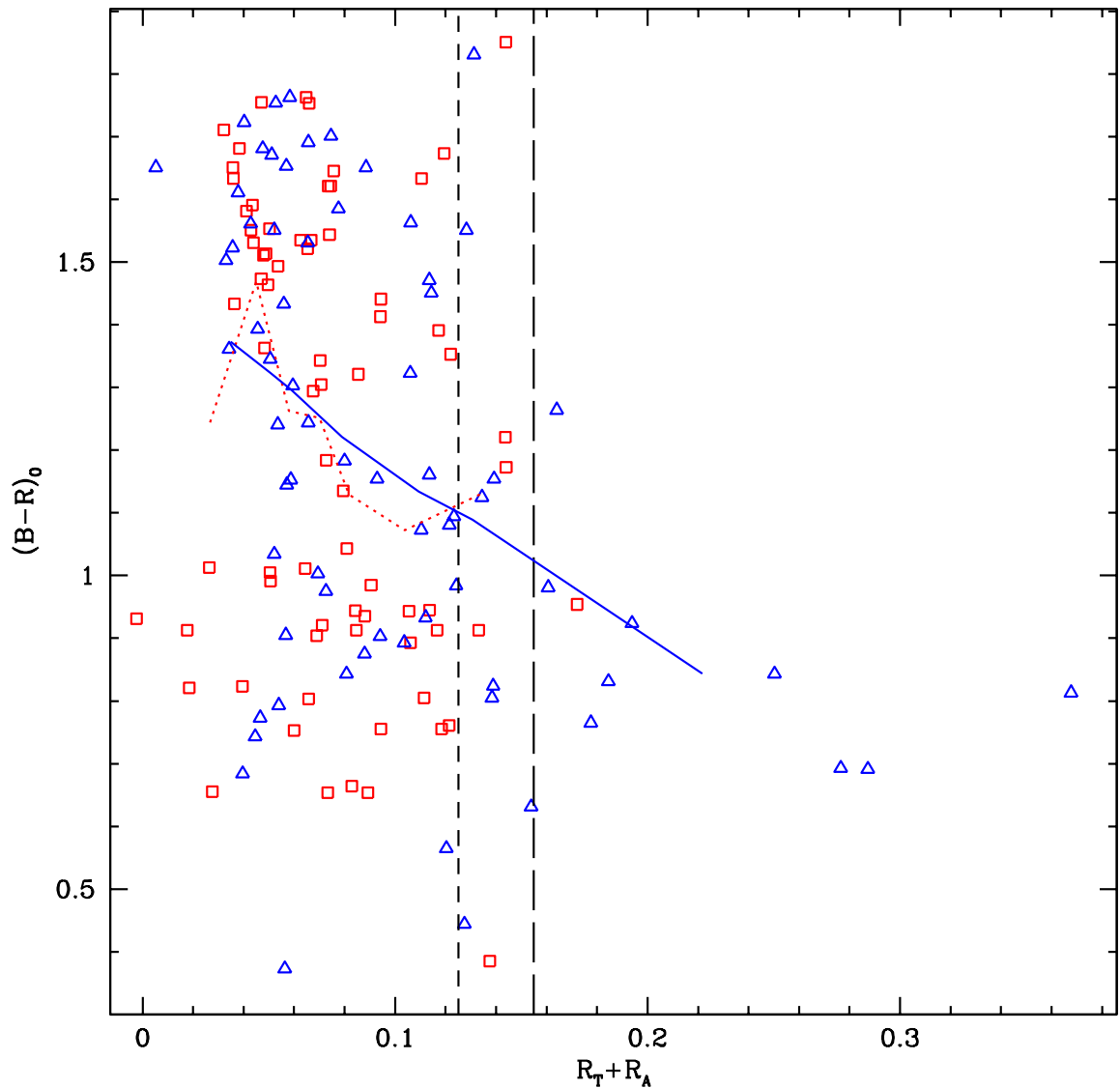}
\caption{Rest-frame $B-R$ color is plotted versus asymmetry
for paired (blue triangles) and isolated (red squares) galaxies.
Galaxies to the right of the short-dashed line are described as
being ``asymmetric''; those to the right of the long-dashed line are
``strongly asymmetric''. 
The dotted red line gives the mean color of isolated 
galaxies in seven bins of $R_T+R_A$, while the 
solid blue line gives the mean color of paired galaxies.
Paired and isolated galaxies have similar distributions in the 
symmetric regime.  Most of the asymmetric galaxies, including the  
seven most asymmetric galaxies, are blue and found in pairs.
\label{figbr0a}}
\end{figure}

In Figure~\ref{figbr0a}, rest-frame color is plotted versus asymmetry.  
Isolated galaxies exhibit a gradual trend towards larger asymmetry for 
bluer galaxies, consistent with the color-asymmetry relationship 
reported by \citet{con00a} for the Frei et al. sample of (non-merging) 
nearby galaxies.  The mean colors of paired and isolated galaxies are 
statistically equivalent for systems classified as symmetric ($R_T+R_A < 0.13$).
On the other hand, the seven most asymmetric galaxies are all paired, 
and are very blue ($(B-R)_0 < 0.95$).  These galaxies lie in a location 
on the color-asymmetry plot similar to that of the merger/interaction induced 
starburst galaxies of \citet{con00b}.  It appears, therefore, that the 
sample of paired galaxies identified in our study is made up of two 
subsets: symmetric galaxies with colors identical to isolated galaxies, 
and asymmetric galaxies with relatively blue colors which have no 
counterparts in the isolated sample.

\subsection{Emission Line Strengths}

An independent measure of current star formation  
can be obtained from the strength of emission line fluxes.
The primary emission line detected in the CNOC2 survey \citep{cnoc2}
is [O{\small\rm II}]3727\AA.  The rest-frame equivalent width of this line, 
denoted EW([OII]), was measured for most of the galaxies of 
interest in this HST study.  Negative EW([OII] refers to emission.
Strong [OII] emission is generally associated with current 
star formation activity \citep[e.g.,][]{ken98}.

The mean EW([OII]) is more negative for paired galaxies than isolated
galaxies, though the difference is relatively small ($5$\AA).  
A K-S test indicates significant differences between the parent 
distributions.  This difference in cumulative distributions 
appears to originate in the strong emission line regime.  
In particular, for galaxies with 
EW([OII]) $< -25$\AA, paired galaxies outnumber isolated galaxies
two to one, despite a slightly smaller sample of paired galaxies versus
isolated galaxies.  These conclusions are similar to those of 
\citet{LK}, who compared the spectrophotometric properties of 
merging and isolated galaxies at low redshift.

\begin{figure}
\epsscale{1.0}
\plotone{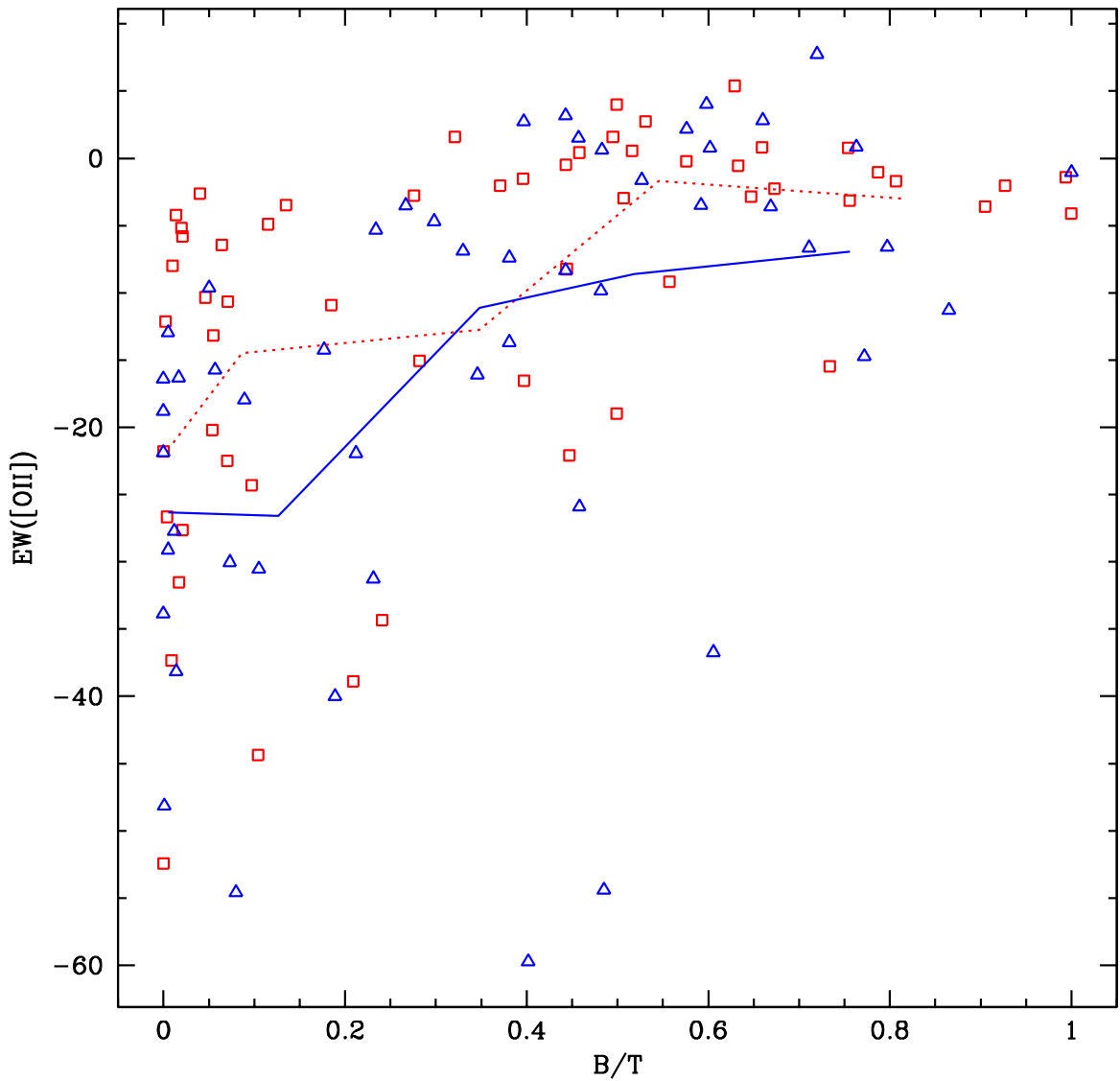}
\caption{[OII]3727 equivalent width is plotted versus bulge fraction 
for paired (blue triangles) and isolated (red squares) galaxies.
The dotted red line gives the mean EW([OII] of isolated 
galaxies in five bins of $B/T$, and shows an anti-correlation 
between emission strength and bulge fraction.  
The solid blue line gives the mean EW([OII] of paired galaxies, 
and exhibits a similar trend.
Paired galaxies exhibit stronger emission than isolated galaxies 
at nearly all bulge fractions.
\label{figeqwbf}}
\end{figure}

\begin{figure}
\epsscale{1.0}
\plotone{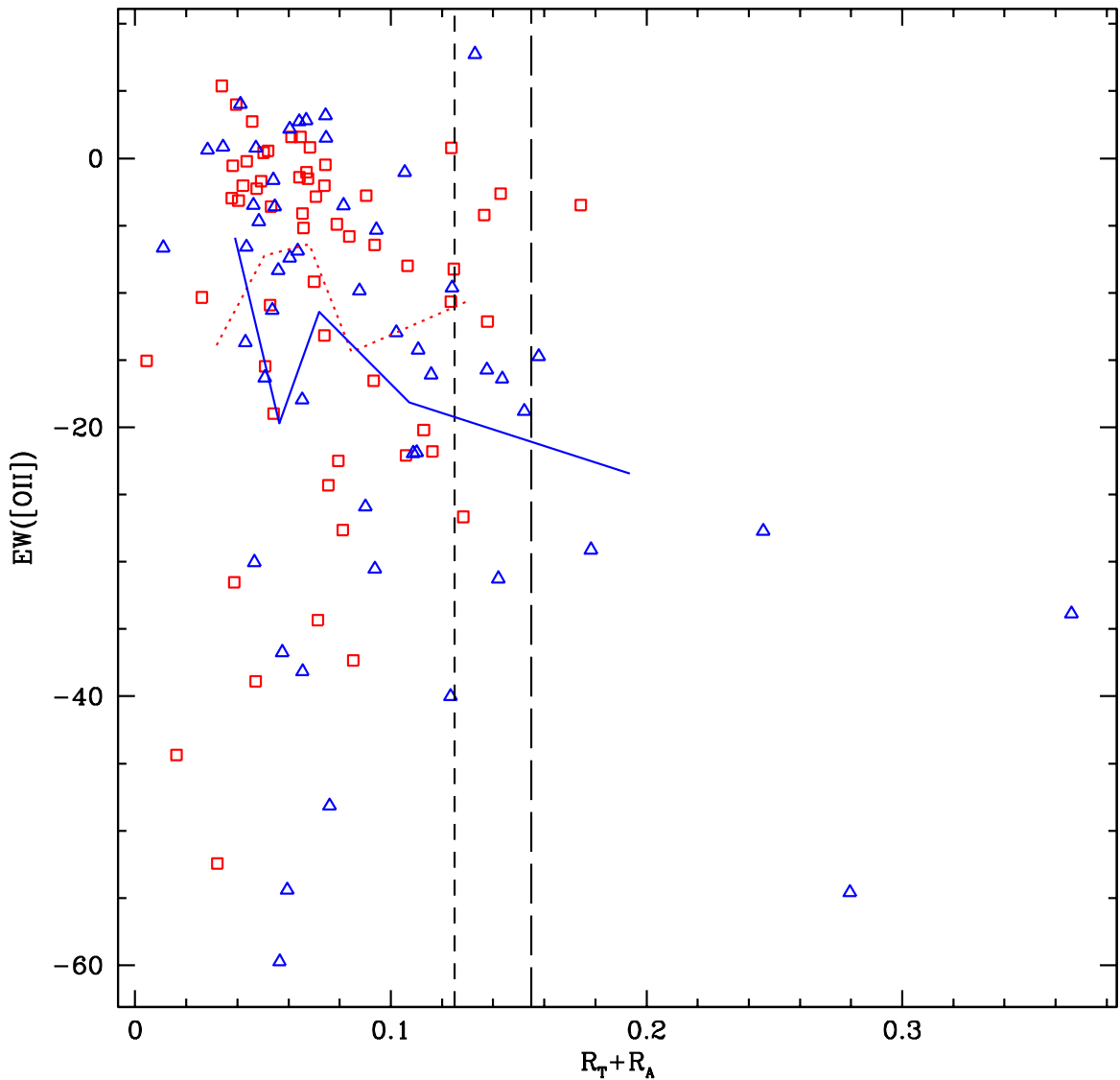}
\caption{[OII]3727 equivalent width is plotted versus asymmetry
for paired (blue triangles) and isolated (red squares) galaxies.
Galaxies to the right of the short-dashed line are described as
being ``asymmetric''; those to the right of the long-dashed line are
``strongly asymmetric''.  
The dotted red line gives the mean EW([OII]) of isolated 
galaxies in five bins of $R_T+R_A$, while 
the solid blue line gives the mean EW([OII]) of paired galaxies.
Paired and isolated galaxies have similar distributions in the 
symmetric regime.  Most of the asymmetric galaxies have strong 
emission and are found in pairs.
\label{figeqwa}}
\end{figure}

In order to see how emission line strength relates to structural parameters, 
we plot EW([OII]) versus bulge fraction (Figure~\ref{figeqwbf}) and 
asymmetry (Figure~\ref{figeqwa}), for both paired and 
isolated galaxies.  While isolated and paired galaxies both exhibit a
general trend towards stronger [OII] emission with smaller bulge fraction, 
paired galaxies are seen to have emission that is stronger than 
isolated galaxies of the same bulge fraction.
Figure~\ref{figeqwa} shows that the subset of paired galaxies 
which are asymmetric exhibit strong [OII] emission.
In addition, direct comparison of colors 
and emission lines for individual galaxies reveals that, for the 
sample of 8 strongly asymmetric paired galaxies in Figure~\ref{figbr0a}, 
4/4 galaxies with emission line measurements have strong emission 
(EW([OII]) $< -27$\AA).  
These results are broadly consistent with the conclusions
of the previous section.

\section{Discussion}\label{discussion}

\subsection{Does the Observed Asymmetry Fraction Make Sense?}

In \S~\ref{asymfrac}, we estimated that roughly 40\% of galaxies 
in merging pairs are asymmetric, and 25\% are strongly asymmetric.  
While these fractions are much higher than the equivalent measurements 
for isolated galaxies, it is worth investigating if these results
are consistent with our hypothesis that the galaxies in our ``merging pair'' 
sample are in fact undergoing mergers at the present time.  
If true, what fraction of these galaxies would be expected to 
exhibit significant asymmetries?  
There are several reasons why this fraction is likely to be substantially 
less than 100\%.  First, even if these 
pairs are likely to merge soon, some may be too early in the
merging process for strong asymmetries to have been produced.   
Also, some galaxies (early types in particular) are less likely to 
exhibit signs of asymmetries than others.  
Finally, the details of the orbits and galaxy rotations affect the 
resulting asymmetries.  It is therefore difficult to predict in advance 
what the asymmetry fraction should be.  

We can, however, compare with other observations.  \citet{con03b}
studied a sample of 66 ultraluminous infrared galaxies (ULIRG's), 
whose infrared emissions are thought to have been triggered
by strong interactions or mergers of gas-rich
spiral galaxies \citep{sm96}.  Their sample would be expected to exhibit 
stronger asymmetries than virtually any other sample of galaxies, including 
the close pairs targeted in this study. 
They found that approximately 50\% of ULIRGs have asymmetries 
consistent with being involved in ongoing major mergers.  They speculate 
that the remaining 50\% of their sample is composed of systems that are 
in either the beginning or the end stages of the merger event, 
when asymmetry is not expected to be strong.  Given this result, 
we conclude that our asymmetry fractions for close galaxy pairs 
are broadly consistent with the hypothesis of P2002 that 
approximately 50\% of the dynamical pairs identified in this study 
are currently undergoing mergers.

\subsection{Enhanced Star Formation}\label{enhanced}

One of the primary goals of modern day cosmology is the determination 
of the star formation history of the universe.  Numerous studies, 
beginning with \citet{madau} and \citet{lilly}, have shown that the 
cosmic star formation rate has decreased by an order of 
magnitude since $z \sim 1$.
A compilation by Hogg (2001)\nocite{hogg} indicates that 
the star formation rate evolves as $(1+z)^{2.7 \pm 0.7}$ over this
redshift range.
It is as yet unclear what is driving this change in star formation rate.
One obvious candidate is galaxy mergers, since mergers are known 
to trigger the formation of new stars in at least some merging systems.
Moreover, the merger rate appears to increase with redshift, at a 
level similar to the increase in the cosmic star formation rate 
\citep[e.g.,][]{cfrs,cnoc2mr}.

At low redshift, there have been several recent studies in which
a clear enhancement in star formation is seen in pairs separated
by less than 20 $h^{-1}$ kpc \citep{barton00,2dfpairs,2dfpairsb,nikolic04}.
At higher redshifts, the results have been less conclusive, primarily 
due to the increased difficulty of identifying dynamically close
galaxy pairs (e.g., Zepf \& Koo 1989\nocite{ZK}; 
Carlberg, Pritchet, \& Infante 1994\nocite{CPI}).  
Using a sample of thirteen dynamical pairs at 
$z \sim 0.3$, \citet{cnoc1mr} found that four of these pairs contain
galaxies with strong [OII] emission, very blue colors, and low 
relative velocities.  While their ground-based images were of marginal
quality, some indications of morphological asymmetry were seen in 
these four pairs.  

In \S~\ref{sf}, we found that paired galaxies have different 
distributions of rest-frame color and emission line strengths 
than isolated galaxies.  In particular, the most asymmetric galaxies have 
unusually blue colors and strong emission lines, and are found almost
exclusively in pairs.  These results are consistent with a scenario
in which merging has induced star formation in a subset of the 
pair sample.  
However, many of the remaining paired galaxies have colors and
emission line strengths consistent with isolated galaxies.
This is hardly surprising, given that the timescale of 
induced star formation is likely to be considerably shorter than
the merging timescale \citep[cf.][]{gillespie03}.  Moreover, 
as outlined in \S~\ref{asymfrac}, our sample of dynamical pairs 
is likely to be a roughly equal mix of merging and non-merging pairs.
As our sample of isolated galaxies contains very few asymmetric galaxies, 
we infer that most of the asymmetric galaxies in the paired sample are 
found in merging pairs.

We interpret this as evidence of induced star formation occurring in 
a significant fraction of merging galaxies at $0.1 < z < 0.6$.
We have arrived at this conclusion using
a sample containing three times as 
many dynamical pairs as the \citet{cnoc1mr} sample, 
and for which we have made quantitative 
measurements of asymmetry using high resolution HST imaging.

\subsection{The Distribution of Bulge Fraction in Paired Galaxies}

In \S~\ref{bulgefraction}, we demonstrated that the bulge fraction 
distributions of paired and isolated galaxies are identical.
One possible interpretation of this result is that 
galaxies of all morphological types have equal
probabilities of being involved in these early-stage mergers, and that
their bulge fractions are not significantly altered by the interaction
until later in the merging process.  
However, if these early stage mergers have triggered star formation, 
as we have argued, one might expect
this extra star formation to occur primarily in the nuclei, 
as predicted by simulations (e.g., Barnes \& Hernquist 1996\nocite{barnes96}).
If so, some paired galaxies might be expected to have boosted $B/T$ 
values, since bulge+disk decompositions cannot 
easily distinguish between a bona fide bulge and a nuclear starburst. 
We would therefore have to conclude that, on average, paired galaxies 
have lower bulge fractions than isolated galaxies.  
We are unable to distinguish between these two 
scenarios with these data.  However, in a study of the nuclear morphology 
in the Toomre sequence of merging galaxies, \citet{laine03} find no evidence
of elevated nuclear emission, except in the latest stage merging systems.  
This implies that nuclear star formation may not be a significant 
factor in our pair sample.

\section{Conclusions}\label{conclusions}

Using HST imaging of CNOC2 galaxies at $0.1 < z < 0.6$, 
we have found clear differences in the structural properties of 
paired and isolated galaxies.  While the distribution of bulge 
fractions are found to be indistinguishable, striking differences are
seen in their asymmetries.  
Assuming that half of our dynamical pairs are 
unlikely to be close enough to merge, we estimate that 
40\% $\pm$ 11\% of merging galaxies are asymmetric, compared with 
9\% $\pm$ 3\% of isolated galaxies. 
This difference is even more pronounced when considering strongly 
asymmetric galaxies (25\% $\pm$ 8\% for merging galaxies versus 
1\% $\pm$ 1\% for isolated galaxies).
These results imply that many of the 
dynamically close galaxy pairs identified by P2002 are
in fact merging, and provide further support for the use of 
close dynamical pairs in estimating the galaxy merger rate and 
its evolution with redshift.  Moreover, these results demonstrate for 
the first time that there is considerable overlap between the 
two most common methods of estimating the galaxy merger rate:  
namely, morphological merger fractions and close pair statistics.

We also find clear evidence for a connection between asymmetry 
and star forming properties in galaxy pairs at intermediate redshift.
Strongly asymmetric paired galaxies are found to be very blue, 
with rest-frame colors in the range $0.65 < (B-R)_0 < 0.95$, 
compared with a mean color of $(B-R)_0 = 1.24$ for all paired galaxies.
In addition, asymmetric galaxies in pairs have 
strong [O{\small\rm II}]3727\AA~emission lines.
Within our sample of 35 galaxy pairs, 
15 pairs contain at least one asymmetric galaxy, and at least 70\% of 
the asymmetric galaxies in these pairs show signs of enhanced star 
formation.  We conclude that close to half of the 
galaxy pairs in our sample are in the process of merging, 
and that the majority of these merging events are accompanied by 
triggered star formation.  When combined with the observed rise 
in the galaxy merger rate with redshift, the 
detection of enhanced star formation in many of these systems 
implies that mergers are at least partially responsible for 
the rise in the cosmic star formation rate with redshift.

\acknowledgments

We thank the referee for comments which significantly improved this manuscript.
Support for Proposal number 8230 was provided by 
NASA through a grant from the Space Telescope Science Institute, which is 
operated by the Association of Universities for Research in Astronomy, 
Incorporated, under NASA contract NAS5-26555.
Support for JG was provided in part by an operating 
grant from Trent University.  DRP, CJP and RGC acknowledge financial 
support from the Natural Sciences and Engineering Research Council of 
Canada.

\clearpage

\clearpage

\clearpage

\clearpage

\clearpage

\clearpage

\clearpage

\clearpage

\clearpage

\clearpage

\clearpage

\clearpage

\clearpage

\clearpage

\clearpage

\begin{table}
\begin{center}
\small
\caption{Properties of Galaxy Pairs.
\label{tabpairprop}}
\begin{tabular}{cccccc}
\tableline\tableline
HST&Mean &$\theta$&$r_p$&$\Delta v$&$\Delta M_R$\\
Target&Redshift&(arcsec)&($h^{-1}$ kpc)&(km/s)&(mag)\\
\tableline
01&0.13339&11.5&17.9&103&2.2\\
04&0.26693&5.1&13.0&232&1.8\\
05&0.47185&2.2&7.7&20&2.0\\
06&0.35090&4.0&12.0&133&1.2\\
07&0.27009&2.5&6.4&319&0.2\\
09&0.38407&2.0&6.2&63&1.3\\
10&0.29829&2.1&5.8&162&1.5\\
11&0.27099&7.0&18.0&189&0.8\\
13&0.29901&7.0&19.2&201&0.4\\
14&0.40769&3.2&10.4&264&1.6\\
15&0.30960&1.3&3.6&82&0.4\\
16&0.14997&6.4&11.0&76&0.4\\
20&0.39666&1.8&5.8&101&0.6\\
25&0.19171&8.6&17.7&423&2.4\\
28&0.24645&2.1&5.1&284&2.2\\
29&0.13640&7.6&12.1&193&0.6\\
30&0.19029&6.8&13.8&360&0.4\\
31&0.39017&1.7&5.5&50&1.4\\
34&0.32471&2.4&7.0&29&0.6\\
35&0.20435&9.2&19.8&50&0.0\\
36&0.39208&2.4&7.6&198&0.0\\
38&0.38954&6.0&18.9&179&0.9\\
39&0.42721&3.8&12.5&78&1.0\\
40&0.55648&5.5&20.5&112&0.4\\
45&0.14608&5.6&9.4&5&2.4\\
46&0.26193&6.9&17.4&404&2.1\\
49&0.26965&2.9&7.5&172&1.7\\
51&0.34926&1.9&5.6&138&0.9\\
52&0.36420&5.0&15.3&171&0.5\\
60&0.51024&3.6&13.0&44&0.1\\
65&0.14744&2.4&4.1&133&0.6\\
67&0.19767&8.4&17.6&48&1.1\\
68&0.40839&4.7&15.2&143&1.9\\
70&0.29718&7.1&19.4&42&1.0\\
72&0.50631&2.1&7.5&125&0.4\\
\tableline
\end{tabular}
\large
\end{center}
\end{table}

\begin{table}
\begin{center}
\scriptsize
\caption{Catalog of Paired Galaxies.
\label{tabpaired}}
\begin{tabular}{ccccccccccr}
\tableline\tableline
HST& CNOC2&RA&Dec&$z$&$R_C$&$M_R$&$B/T$&$R_T$+$R_A$&$(B-R)_0$&$[OII]$\\
ID&Name&(J2000)&(J2000)&&&&&&&E.W.\\
\tableline
01A&0223-100810&02:26:53.50&00:02:48.0&0.13358&17.0&-21.2&0.69&0.04&1.52&------\\
01B&0223-100778&02:26:53.07&00:02:38.6&0.13319&19.2&-19.0&0.67&0.08&1.18&------\\
04A&0223-140091&02:25:22.94&00:07:12.4&0.26742&18.8&-20.9&0.01&0.25&0.84&-27.7\\
04B&0223-140075&02:25:23.20&00:07:09.2&0.26644&20.6&-19.1&0.46&0.09&0.90&-25.9\\
05A&0223-062160&02:25:50.10&01:00:41.5&0.47190&18.8&-22.9&0.58&0.06&1.76&2.2\\
05B&0223-062166&02:25:50.25&01:00:42.0&0.47180&20.1&-20.9&0.40&0.06&2.81&2.7\\
06A&0223-130708&02:25:09.06&00:18:05.2&0.35060&19.0&-21.7&0.48&0.03&1.50&0.6\\
06B&0223-130707&02:25:09.33&00:18:04.8&0.35120&20.1&-20.5&0.38&0.04&1.72&-13.7\\
07A&0223-191878&02:23:28.45&-00:03:55.1&0.27077&19.2&-20.5&0.00&0.11&0.93&-21.9\\
07B&0223-191879&02:23:28.29&-00:03:55.0&0.26942&19.6&-20.2&0.01&0.10&0.89&-13.0\\
09A&0223-121099&02:26:17.22&00:12:46.9&0.38393&19.4&-21.5&0.33&0.06&1.65&-6.9\\
09B&0223-121101&02:26:17.09&00:12:47.3&0.38422&20.6&-20.2&0.66&0.07&1.00&2.8\\
10A&0223-030307&02:25:49.16&00:30:33.1&0.29864&19.1&-21.1&0.48&0.06&1.43&-54.4\\
10B&0223-030309&02:25:49.02&00:30:33.7&0.29794&20.4&-19.6&0.00&0.08&0.84&-48.2\\
11A&0223-050675&02:25:55.13&00:47:18.6&0.27059&19.6&-20.2&0.18&0.11&1.07&-14.2\\
11B&0223-050679&02:25:55.59&00:47:18.9&0.27139&20.5&-19.4&0.38&0.06&1.15&-7.4\\
13A&0223-030266&02:25:49.87&00:30:23.7&0.29945&19.5&-20.7&0.21&0.11&1.32&-21.9\\
13B&0223-030259&02:25:49.42&00:30:21.9&0.29858&19.9&-20.3&0.44&0.06&1.30&-8.3\\
14A&0223-031099&02:25:48.86&00:34:00.2&0.40831&19.6&-20.9&0.08&0.28&0.69&-54.6\\
14B&0223-031118&02:25:48.90&00:34:03.4&0.40707&21.1&-19.3&0.40&0.06&0.37&-59.7\\
15A&0223-131144&02:25:25.89&00:20:06.4&0.30978&20.0&-20.4&0.59&0.05&1.39&-3.5\\
15B&0223-131141&02:25:25.82&00:20:05.8&0.30942&20.2&-20.0&1.00&0.11&1.56&-1.0\\
16A&0223-160751&02:25:02.42&00:01:13.1&0.15011&20.0&-18.3&0.01&0.05&0.77&------\\
16B&0223-160754&02:25:02.84&00:01:14.4&0.14982&20.4&-17.9&0.32&0.04&0.74&------\\
20A&0223-040345&02:26:13.19&00:38:25.3&0.39643&20.1&-20.4&0.00&0.37&0.81&-33.9\\
20B&0223-040350&02:26:13.24&00:38:26.9&0.39690&20.7&-19.9&0.07&0.05&0.79&-30.0\\
25A&0920-010722&09:24:03.22&37:04:52.4&0.19087&17.1&-22.0&1.00&0.07&1.70&------\\
25B&0920-010688&09:24:03.32&37:04:43.9&0.19255&19.5&-19.6&0.00&0.13&1.55&------\\
28A&0920-081315&09:24:39.40&37:08:15.3&0.24586&18.2&-21.6&0.67&0.05&1.55&-3.6\\
28B&0920-081301&09:24:39.46&37:08:13.4&0.24704&20.4&-19.3&0.48&0.09&1.65&-9.8\\
29A&0920-050264&09:23:43.09&37:31:42.9&0.13676&18.6&-19.7&0.86&0.11&1.47&------\\
29B&0920-050258&09:23:42.49&37:31:40.2&0.13603&19.0&-19.1&0.00&0.29&0.69&------\\
30A&0920-010879&09:24:08.24&37:05:42.5&0.18958&18.8&-20.3&0.73&0.11&1.45&------\\
30B&0920-010860&09:24:07.95&37:05:36.8&0.19101&19.2&-19.9&0.80&0.04&1.56&------\\
31A&0920-131238&09:23:12.63&37:07:04.7&0.39028&18.9&-22.1&0.53&0.05&1.67&-1.6\\
31B&0920-131230&09:23:12.50&37:07:04.0&0.39005&20.4&-20.7&0.46&0.07&1.69&1.5\\
\tableline
\end{tabular}
\large
\end{center}
\end{table}

\addtocounter{table}{-1}
\begin{table}
\begin{center}
\scriptsize
\caption{Catalog of Paired Galaxies -- {\it continued.}
}
\begin{tabular}{ccccccccccr}
\tableline\tableline
HST&CNOC2&RA&Dec&$z$&$R_C$&$M_R$&$B/T$&$R_T$+$R_A$&$(B-R)_0$&$[OII]$\\
ID&Name&(J2000)&(J2000)&&&&&&&E.W.\\
\tableline
34A&0920-020557&09:23:30.71&37:11:19.6&0.32477&19.6&-20.6&0.01&0.18&0.83&-29.1\\
34B&0920-020566&09:23:30.87&37:11:21.0&0.32464&20.1&-20.0&0.00&0.15&0.63&-18.8\\
35A&0920-020237&09:24:06.44&37:09:45.3&0.20425&19.7&-19.4&0.16&0.16&0.98&------\\
35B&0920-020209&09:24:05.90&37:09:38.8&0.20445&19.7&-19.4&0.02&0.11&1.16&------\\
36A&0920-160826&09:22:27.87&36:46:51.9&0.39162&19.9&-21.0&0.30&0.05&1.24&-4.7\\
36B&0920-160832&09:22:27.83&36:46:54.2&0.39254&20.0&-21.0&0.80&0.04&1.61&-6.6\\
38A&0920-150212&09:23:13.80&36:45:29.3&0.38996&20.2&-20.9&0.60&0.05&1.68&0.8\\
38B&0920-150213&09:23:14.30&36:45:29.5&0.38913&21.0&-20.0&0.76&0.03&1.36&0.8\\
39A&0920-151065&09:23:00.43&36:49:45.3&0.42740&20.3&-21.1&0.72&0.13&1.83&7.8\\
39B&0920-151064&09:23:00.12&36:49:45.2&0.42703&21.2&-20.1&0.71&0.01&1.65&-6.6\\
40A&0920-180867&09:21:42.11&36:37:57.4&0.55677&20.7&-21.2&0.35&0.12&1.08&-16.1\\
40B&0920-180845&09:21:42.49&36:37:54.4&0.55619&20.7&-21.6&0.44&0.07&1.53&3.2\\
45A&1447-101867&14:50:25.13&08:56:28.0&0.14609&18.1&-20.2&0.13&0.19&0.92&------\\
45B&1447-101865&14:50:25.48&08:56:25.8&0.14607&20.4&-17.8&0.00&0.13&0.44&------\\
46A&1447-140046&14:48:55.81&08:56:44.4&0.26108&18.2&-21.6&0.05&0.12&1.09&-9.6\\
46B&1447-140054&14:48:56.27&08:56:45.7&0.26278&20.3&-19.4&0.02&0.05&1.03&-16.3\\
49A&1447-110843&14:49:26.91&08:52:11.1&0.27002&19.1&-20.7&0.05&0.12&0.98&------\\
49B&1447-110838&14:49:27.10&08:52:10.2&0.26929&20.8&-19.0&0.00&0.14&1.15&-16.4\\
51A&1447-051065&14:49:55.08&09:38:44.9&0.34895&19.2&-21.5&0.86&0.05&1.75&-11.3\\
51B&1447-051056&14:49:54.98&09:38:43.7&0.34957&20.0&-20.6&0.77&0.16&1.26&-14.7\\
52A&1447-082155&14:50:05.66&09:11:52.5&0.36381&19.3&-21.1&0.06&0.14&0.82&-15.7\\
52B&1447-082172&14:50:05.87&09:11:56.3&0.36459&20.1&-20.6&0.23&0.09&1.15&-5.3\\
60A&1447-040704&14:49:38.98&09:29:31.0&0.51035&20.9&-20.6&0.60&0.04&0.68&4.0\\
60B&1447-040709&14:49:38.74&09:29:31.0&0.51013&21.3&-20.4&0.09&0.07&1.24&-18.0\\
65A&2148-042181&21:51:24.69&-05:07:10.2&0.14719&18.8&-19.7&0.56&0.05&1.34&------\\
65B&2148-042170&21:51:24.76&-05:07:12.4&0.14770&19.3&-19.1&0.08&0.13&1.12&------\\
67A&2148-031553&21:51:12.48&-05:17:42.8&0.19758&19.1&-19.9&0.69&0.18&0.77&------\\
67B&2148-031529&21:51:12.88&-05:17:48.6&0.19777&20.2&-18.7&0.04&0.06&0.91&------\\
68A&2148-071349&21:51:22.15&-04:45:14.3&0.40806&19.5&-21.6&0.27&0.08&1.58&-3.5\\
68B&2148-071332&21:51:22.09&-04:45:18.9&0.40873&21.1&-19.7&0.01&0.07&0.97&-38.1\\
70A&2148-051930&21:51:26.12&-04:58:16.6&0.29727&20.0&-19.9&0.10&0.09&0.88&-30.5\\
70B&2148-051898&21:51:25.97&-04:58:23.3&0.29709&21.1&-18.9&0.61&0.06&1.15&-36.7\\
72A&2148-162261&21:50:16.57&-05:45:45.8&0.50600&20.8&-20.4&0.19&0.12&0.56&-40.0\\
72B&2148-162264&21:50:16.46&-05:45:44.4&0.50663&21.2&-20.0&0.23&0.14&0.81&-31.2\\
\tableline
\end{tabular}
\large
\end{center}
\end{table}

\begin{table}
\begin{center}
\scriptsize
\caption{Catalog of Isolated Galaxies.
\label{tabiso}}
\begin{tabular}{ccccccccccr}
\tableline\tableline
HST& CNOC2&RA&Dec&$z$&$R_C$&$M_R$&$B/T$&$R_T$+$R_A$&$(B-R)_0$&$[OII]$\\
ID&Name&(J2000)&(J2000)&&&&&&&E.W.\\
\tableline
01&0223-100974&02:26:48.44&00:03:40.0&0.57410&21.0&-20.9&0.02&0.07&0.80&-5.2\\
02&0223-100908&02:26:47.44&00:03:19.9&0.16531&21.2&-17.4&0.12&0.06&0.75&------\\
03&0223-100681&02:26:52.15&00:02:12.2&0.23069&19.2&-20.4&0.56&0.07&1.34&-9.2\\
04&0223-020713&02:25:50.07&00:25:12.0&0.37752&20.5&-20.4&0.28&0.09&1.41&-2.8\\
05&0223-140376&02:25:25.74&00:08:32.8&0.28505&20.0&-20.1&0.50&0.05&1.36&-19.0\\
06&0223-140073&02:25:15.60&00:07:06.9&0.61474&20.8&-21.0&0.00&0.13&0.91&-26.7\\
07&0223-062315&02:25:52.53&01:01:24.2&0.30465&19.2&-21.1&0.91&0.05&1.55&-3.6\\
08&0223-070252&02:25:52.40&01:02:17.8&0.30533&19.8&-20.4&0.45&0.12&1.35&------\\
09&0223-130667&02:25:12.65&00:17:54.5&0.35944&18.8&-21.9&0.46&0.05&1.49&0.4\\
10&0223-130491&02:25:12.74&00:17:06.4&0.29227&19.1&-21.0&0.52&0.05&1.47&0.6\\
11&0223-130656&02:25:13.25&00:17:51.7&0.30653&19.4&-20.9&0.75&0.12&1.67&0.8\\
12&0223-130882&02:25:11.73&00:18:50.4&0.39001&20.4&-20.7&0.79&0.07&1.75&-1.0\\
13&0223-191602&02:23:25.41&-00:04:49.1&0.24321&20.2&-19.3&0.07&0.08&1.04&-22.5\\
14&0223-191492&02:23:25.12&-00:05:07.9&0.27001&20.5&-19.5&0.45&0.11&1.63&-22.1\\
15&0223-030382&02:25:51.62&00:30:51.3&0.29849&19.6&-20.6&0.73&0.05&1.51&-15.4\\
16&0223-021797&02:25:47.12&00:29:26.3&0.36418&19.2&-21.6&0.76&0.04&1.63&-3.1\\
17&0223-030996&02:25:45.61&00:33:33.4&0.30046&21.0&-18.9&0.02&0.04&0.82&-31.5\\
18&0223-131066&02:25:31.36&00:19:42.4&0.36261&20.3&-20.2&0.05&0.03&1.01&-10.3\\
19&0223-161138&02:25:06.88&00:02:32.8&0.33161&20.3&-20.2&0.50&0.04&1.43&4.0\\
20&0223-040234&02:26:19.19&00:37:56.7&0.26701&19.7&-20.1&0.01&0.11&0.94&-8.0\\
21&0223-040342&02:26:11.95&00:38:24.1&0.18752&18.5&-20.4&0.03&0.11&0.89&------\\
22&0223-031673&02:26:05.17&00:36:25.5&0.42483&21.2&-19.7&0.00&0.12&0.91&-21.8\\
23&0223-171157&02:24:53.29&-00:06:20.5&0.38609&20.6&-20.0&0.10&0.08&0.91&-24.3\\
24&0223-170849&02:24:50.90&-00:07:12.7&0.39019&21.2&-19.4&0.10&0.02&0.91&-44.4\\
25&0223-170801&02:24:47.75&-00:07:21.9&0.29572&19.6&-20.5&0.01&0.14&1.17&-4.2\\
26&0223-171287&02:24:42.76&-00:05:57.8&0.55464&20.5&-21.8&0.49&0.06&1.76&1.6\\
27&0920-010390&09:24:05.97&37:03:27.6&0.37363&19.6&-21.4&0.51&0.04&1.68&-2.9\\
28&0920-010331&09:24:06.23&37:03:09.7&0.37234&20.3&-20.6&0.63&0.03&1.71&5.4\\
29&0920-010500&09:24:03.30&37:03:53.7&0.19309&20.6&-18.2&0.06&0.02&0.82&------\\
30&0920-010748&09:24:04.19&37:04:59.4&0.20159&18.4&-20.8&0.57&0.04&1.59&------\\
31&0920-080744&09:24:10.53&37:05:24.4&0.32611&19.8&-20.8&0.44&0.12&1.39&-8.2\\
32&0920-081152&09:24:40.62&37:07:32.7&0.24619&20.4&-19.4&0.04&0.11&2.05&------\\
33&0920-050330&09:23:43.87&37:32:02.2&0.24630&19.4&-20.3&0.44&0.09&1.44&------\\
34&0920-050346&09:23:52.14&37:32:07.0&0.32093&20.6&-19.8&0.53&0.04&1.55&------\\
35&0920-011194&09:24:04.30&37:07:04.7&0.23134&18.7&-20.9&0.63&0.04&1.58&-0.6\\
36&0920-011167&09:24:08.27&37:06:55.3&0.37291&19.0&-21.9&0.93&0.04&1.65&-2.0\\
37&0920-131204&09:23:13.70&37:06:56.7&0.39211&18.8&-22.2&0.66&0.07&1.62&0.8\\
38&0920-131311&09:23:20.06&37:07:21.6&0.24337&18.3&-21.4&0.58&0.04&1.53&-0.2\\
39&0920-131351&09:23:21.04&37:07:31.1&0.24503&20.1&-19.5&0.21&0.05&0.99&-38.9\\
\tableline
\end{tabular}
\large
\end{center}
\end{table}

\addtocounter{table}{-1}
\begin{table}
\begin{center}
\scriptsize
\caption{Catalog of Isolated Galaxies -- {\it continued.}
}
\begin{tabular}{ccccccccccr}
\tableline\tableline
HST& CNOC2&RA&Dec&$z$&$R_C$&$M_R$&$B/T$&$R_T$+$R_A$&$(B-R)_0$&$[OII]$\\
ID&Name&(J2000)&(J2000)&&&&&&&E.W.\\
\tableline
40&0920-020409&09:23:33.83&37:10:39.4&0.24740&18.1&-21.6&0.44&0.07&1.62&-0.5\\
41&0920-020268&09:23:31.84&37:09:55.6&0.24656&18.4&-21.1&0.04&0.14&1.22&-2.6\\
42&0920-081330&09:24:12.56&37:08:22.9&0.46150&19.9&-21.7&0.32&0.06&2.20&1.6\\
43&0920-020299&09:24:07.39&37:10:01.3&0.24655&20.0&-19.7&0.53&0.05&1.51&2.7\\
44&0920-160943&09:22:34.01&36:47:14.5&0.11019&17.4&-20.3&0.61&0.14&2.22&------\\
45&0920-150040&09:23:13.93&36:44:43.1&0.44119&20.8&-20.3&0.28&0.00&0.93&-15.1\\
46&0920-150181&09:23:15.54&36:45:20.7&0.18807&19.6&-19.3&0.03&0.06&1.01&------\\
47&0920-150193&09:23:23.48&36:45:25.5&0.55982&20.3&-21.4&0.06&0.07&0.92&-13.2\\
48&0920-151228&09:22:56.31&36:50:33.4&0.42814&20.8&-20.5&0.65&0.07&1.52&-2.9\\
49&0920-180260&09:21:45.07&36:36:29.9&0.46012&20.8&-20.3&0.07&0.12&0.76&-10.7\\
50&0920-181107&09:21:44.43&36:38:30.9&0.43685&20.0&-21.2&0.06&0.09&1.32&-6.4\\
51&0920-030026&09:23:40.18&37:16:05.2&0.32032&20.7&-19.4&0.00&0.14&1.85&-12.2\\
52&1447-101860&14:50:24.79&08:56:23.9&0.37270&20.4&-20.4&0.40&0.07&1.29&-1.5\\
53&1447-151214&14:48:53.47&08:55:20.0&0.51551&21.3&-19.8&0.07&0.07&0.65&------\\
54&1447-110526&14:49:25.59&08:50:52.9&0.60697&19.7&-22.2&0.14&0.17&0.95&-3.5\\
55&1447-110482&14:49:28.21&08:50:37.6&0.37317&20.9&-19.6&0.02&0.08&0.94&-27.6\\
56&1447-110595&14:49:29.40&08:51:07.8&0.55393&21.1&-20.5&0.01&0.09&0.65&-37.3\\
57&1447-051337&14:49:52.34&09:39:47.5&0.43315&20.6&-20.6&0.24&0.07&1.18&-34.3\\
58&1447-082395&14:50:01.48&09:12:43.2&0.30347&18.9&-21.5&1.00&0.07&1.54&-4.1\\
59&1447-040758&14:49:34.82&09:29:43.1&0.13114&20.1&-18.0&0.28&0.07&0.90&------\\
60&1447-040497&14:49:38.34&09:28:34.9&0.40724&20.1&-20.7&0.02&0.08&0.66&-5.8\\
61&1447-040441&14:49:40.07&09:28:23.0&0.46205&20.6&-20.9&0.67&0.05&1.46&-2.3\\
62&2148-070517&21:51:09.61&-04:47:50.4&0.15444&17.8&-20.8&0.80&0.08&1.65&------\\
63&2148-191639&21:48:47.50&-05:57:32.6&0.25759&19.2&-20.4&0.40&0.09&0.98&-16.6\\
64&2148-191186&21:48:47.92&-05:59:02.3&0.15252&19.8&-18.5&0.00&0.09&0.76&------\\
65&2148-191213&21:48:47.63&-05:58:56.9&0.55673&21.1&-21.0&0.81&0.09&2.04&------\\
66&2148-042309&21:51:19.50&-05:06:47.8&0.54593&21.5&-19.8&0.00&0.03&0.66&-52.4\\
67&2148-042026&21:51:25.10&-05:07:42.4&0.14776&17.3&-21.1&0.43&0.07&1.53&------\\
68&2148-031234&21:51:12.94&-05:18:43.6&0.73200&20.2&-21.7&0.32&0.14&0.39&------\\
69&2148-071573&21:51:20.96&-04:44:38.2&0.28803&19.7&-20.4&0.99&0.06&1.53&-1.4\\
70&2148-071141&21:51:18.04&-04:45:55.9&0.12729&17.7&-20.3&0.04&0.09&0.94&------\\
71&2148-071059&21:51:18.02&-04:46:09.0&0.28720&18.3&-21.8&0.81&0.05&1.75&-1.7\\
72&2148-070881&21:51:20.76&-04:46:38.5&0.53628&20.9&-20.7&0.18&0.05&1.00&-10.9\\
73&2148-051766&21:51:31.74&-04:59:00.3&0.15642&20.5&-17.9&0.09&0.12&0.76&------\\
74&2148-161891&21:50:19.89&-05:47:13.0&0.14443&19.8&-18.5&0.61&0.11&0.94&------\\
75&2148-162029&21:50:20.47&-05:46:39.3&0.37508&19.9&-20.9&0.37&0.07&1.30&-2.0\\
76&2148-162168&21:50:18.51&-05:46:06.5&0.31699&20.6&-19.7&0.12&0.08&1.14&-4.9\\
77&2148-162322&21:50:19.63&-05:45:30.1&0.39176&21.5&-19.1&0.05&0.11&0.81&-20.2\\
\tableline
\end{tabular}
\large
\end{center}
\end{table}

\begin{table}
\begin{center}
\caption{Comparison of Paired and Isolated Galaxies.
\label{tabprop}}
\begin{tabular}{cccc}
\tableline\tableline
Property&Paired&Isolated&KS Prob (\%)\\
\tableline
N&70&77&\\
$z$&0.31 $\pm$ 0.01& 0.34  $\pm$ 0.02&66\\
$R_C$&19.8 $\pm$ 0.1&19.9 $\pm$ 0.1&44\\
$M_R$&$-20.3 \pm 0.1$&$-20.4 \pm 0.1$&47\\
$(B-R)_0$&1.20 $\pm$ 0.05&1.27 $\pm$ 0.05&66\\ 
$B/T$&0.36 $\pm$ 0.04& 0.35 $\pm$ 0.03&95\\
$R_T+R_A$&0.098 $\pm$ 0.008&0.076 $\pm$ 0.004&16\\
$[OII]EW$&$-15.8 \pm 2.3$&$-10.5 \pm 1.7$&11\\
\tableline
\end{tabular}
\end{center}
\end{table}

\end{document}